\documentclass[letterpaper,12pt]{article}
\pdfoutput=1 
\usepackage{epsfig}
\usepackage{float}
\usepackage{dsfont}
\usepackage{jheppub} 

\usepackage{bbold}

\makeatletter
\def\@fpheader{\relax}
\makeatother

\usepackage{subcaption}
\usepackage{amsmath,amsfonts,amssymb}
\usepackage{psfrag}
\usepackage{enumerate}
\usepackage{mathrsfs}
\usepackage{graphicx}
\usepackage{wrapfig}
\usepackage{xcolor}
\usepackage{caption}
\usepackage{amsthm}

\renewcommand{\thanks}[1]{\footnote{#1}} 

\newcommand{\be}{\begin{equation}}
\newcommand{\bea}{\begin{eqnarray}}
\newcommand{\eea}{\end{eqnarray}}
\newcommand{\beq}{\begin{equation}}
\newcommand{\ee}{\end{equation}}

\newcommand{\myeq}[1]{\begin{equation} #1 \end{equation}}
\newcommand{\myal}[1]{\begin{align} #1 \end{align}}

\def\simleq{\; \raise0.3ex\hbox{$<$\kern-0.75em
\raise-1.1ex\hbox{$\sim$}}\; }
\def\simgeq{\; \raise0.3ex\hbox{$>$\kern-0.75em
\raise-1.1ex\hbox{$\sim$}}\; }

\def\bi{\begin{itemize}}
\def\ei{\end{itemize}}

\def\sc{\setcounter{equation}{0}}

\usepackage{color}

\subheader{}

\title{Inside the Hologram: Reconstructing the bulk observer's experience}

\author[a]{Daniel Louis Jafferis,}   \author[b]{Lampros Lamprou}

\affiliation[a]{Center for the Fundamental Laws of Nature, Harvard University, Cambridge, MA 02138, USA}
\affiliation[b]{Center for Theoretical Physics, Massachusetts Institute of Technology, Cambridge, MA 02139-4307,USA}

\abstract{We develop a holographic framework for describing the experience of bulk observers in AdS/CFT, that allows us to compute the proper time and energy distribution measured along any bulk worldline. Our method is formulated directly in the CFT language and is universal: It does not require knowledge of the bulk geometry as an input. When used to propagate operators along the worldline of an observer falling into an eternal black hole, our proposal resolves a conceptual puzzle raised by Marolf and Wall. Notably, the prescription does not rely on an external dynamical Hamiltonian or the AdS boundary conditions and is, therefore, outlining a general framework for the emergence of time.}

\begin{document}

\maketitle
\flushbottom
\tableofcontents

\section{Inside a quantum system}
\label{sec:intro}
Suppose your adventurous colleague jumped in the AdS black hole you created in your lab's quantum computer by simulating ${\cal N}=4$ Super Yang-Mills. What did their Geiger counter register along their journey and at what age did they meet their inevitable end?

Gauge/gravity duality \cite{Maldacena:1997re} has offered a wealth of insights on the microscopic description of black holes, as observed from an asymptotic frame. These include their entropy, fast scrambling dynamics \cite{Sekino:2008he, Shenker:2013pqa} and unitarity of the Hawking evaporation process \cite{Penington:2019npb, Almheiri:2019psf}. In contrast, the infalling observer's experience remains mysterious, owing to the lack of holographic reconstruction techniques that penetrate bulk horizons. The difficulty in posing, even in principle, operationally meaningful questions such as the amount of time or energy measured by observers behind horizons highlights a gap in our understanding of AdS/CFT: The absence of a CFT framework for describing physics in an \emph{internal reference frame}. 

To catalyze progress in this direction, we pursue an observer-centric approach to bulk reconstruction.\footnote{Related attempts to describe physics from an internal observer's point of view include \cite{Anninos:2011af, vanBreukelen:2017dul}, as well as the series of works \cite{Yoshida:2018ybz, Yoshida:2019qqw, Yoshida:2019kyp, Yoshida:2020wpd} which includes a few interpretational claims that deviate somewhat from the proposal presented in this paper.} Even for observers that do not fall into black holes, no general method for determining how their experience is encoded in the CFT is known, particularly without having to solve the bulk theory directly. This is closely related to the fact that CFT operators are attuned to an external description of the quantum system, while the observer is associated to an internal frame of reference.

Any observer made out of bulk matter is simply a suitable subsystem of the dual Conformal Field Theory\footnote{Our construction does not rely on conformal symmetry. This generality is crucial for it to apply in non-vacuum states with semi-classical bulk duals}. In our work, this observer will be a black hole entangled with an external reference; the subsystem available for their experiments consists of operators within the black hole ``atmosphere'' (Section \ref{sec:observers}). The entangled reference provides an external way to describe the frame associated to the observer. The virtue of such a probe black hole is in providing a particularly simple model of a subsystem, related by a unitary transformation to the thermofield double state \cite{Maldacena:2001kr}. 

The probe black hole will be introduced near the boundary, and then allowed to propagate under time evolution before returning to our possession at a later time\footnote{This precludes exploring behind horizons in the particular setup of this work. Nevertheless, our framework contains lessons for black hole interior reconstruction which we discuss in Section \ref{sec:discussion}.}. The only assumption about the bulk state we make is that it is described by a semi-classical spacetime, whose features we wish to probe in the classical limit. In this setup, we start by solving the following problem: Assuming CFT knowledge of the atmosphere degrees of freedom at the initial and final timeslices, and of the CFT Hamiltonian, how much proper time did the observer's clock measure and what energy distribution was detected by their calorimeter?

In Sections \ref{sec:time} and \ref{sec:particles}, we explain how to read off this information from the boundary unitary $V_H(0,t)$ that relates the initial $(t_i=0)$ and final $(t_f=t)$ local atmosphere operators. The result is universal within its domain of validity, and does not require as an input the solution for the bulk spacetime. The key ingredient is the modular Hamiltonian of the black hole, $K=-\log \rho$, defined by its reduced density matrix $\rho$ after tracing out the reference system. In a nutshell, we propose that the decomposition of $\log V_H$ in terms of (approximate) ``eigen-operators'' of $K$ has the schematic form:
\myeq{i\log V_H(0,t) = \frac{\tau(t)}{2\pi} \int d\Omega_{d-2} \, f(\Omega) \,G_{2\pi}(\Omega) + \frac{\tau(t)}{2\pi} K +\text{other zero-modes} + O(e^{-\tau}, N^{-1}) \label{claim1}}
The coefficient $\tau(t)$ of the modular operator is the \emph{proper time}. $G_{2\pi}(\Omega)$ are the modular scrambling modes, satisfying $[K, G_{2\pi}]= -2\pi i G_{2\pi}$ \cite{deBoer:2019uem} thus growing exponentially under modular flow, and the coefficient $f(\Omega)$ depends on the expectation value of the horizon null energy flux $ \int_{x^+=0} dx^- \langle T_{--}\rangle $ in the frame of the moving black hole. A non-vanishing scrambling mode coefficient in (\ref{claim1}) is, therefore, a signature of particle detection along the bulk worldline. The other zero mode contributions describe the precession of the observer's symmetry frame, e.g. a local rotation about the black hole. Our result is valid when the proper time is shorter than the scrambling time of our probe black hole $ \tau(t) \lesssim  \log S_{BH}$.

The basic intuition, from the point of view of the quantum system, is that proper time is measured by the phase, $e^{i m \tau}$, of the state, as in Feynman's path integral for a particle worldline. In our setup, the analog of the rest mass, $m$ is provided by the local energy, conveniently given by the Hamiltonian of the reference system which is exactly AdS Schwarzschild. This corresponds to the modular Hamiltonian for the spacetime being probed. An important additional ingredient is that we keep track of the unitary $V_H(0,t)$ that relates the presumed known {\it operator} algebra of the probe black hole atmosphere at the beginning and end of its journey through the bulk, to correctly translate between the initial and final states. This is necessary to have  a well-defined comparison without requiring the input of a collection of CFT operators associated to translations in a particular bulk time slicing. 

In the bulk language, we use the fact that the modular Hamiltonian acts approximately as the local Schwarzschild time evolution relative to the extremal surface, in a sense that we make more precise below. An important point is that because the total state is a single sided unitary transformation of the thermofield double, the corrections to the geometric action of the modular flow can be regarded as transients due to excitations of the bulk fields near the probe black hole, which in our setup do not affect the initial and final atmosphere operators. The only persistent effect is a shift of the extremal surface relative to the probe black hole causal horizon, with respect to which we wish to define the atmosphere operators. This is encoded by the scrambling modes in eq. (\ref{claim1}) and is explained in detail in Section \ref{sec:particles}.

In Section \ref{subsec:timedilation}, we employ this technology to holographically measure the time dilation witnessed by two twins who embark on separate journeys and compare clocks at their future reunion. In the dual quantum description, the twins select two time-dependent families of modular Hamiltonians, describing the evolution of their state, which are simply related when the siblings meet at the initial and final moments, thus forming a ``closed loop''. The proper time difference experienced by the twins is computed by two \emph{intrinsic} properties of this modular loop: the modular Berry holonomy \cite{Czech:2017zfq, Czech:2018kvg, Czech:2019vih} (Section \ref{subsec:KandW}) and the modular zero mode component of the CFT Hamiltonian integrated along the path.

The experience of an observer falling into a black hole is discussed in Section \ref{sec:discussion}. Our framework naturally resolves a conceptual puzzle raised by Marolf and Wall \cite{Marolf:2012xe} regarding the ability of an observer that enters from one side of an AdS wormhole to receive signals coming from the other, despite the dynamical decoupling of the two exterior regions. Moreover, our proposal outlines an interesting perspective on the ``problem of time'' in quantum gravity in general, an early form of which was anticipated in \cite{Connes:1994hv}. Our eq.~(\ref{claim1})  links the geometric notion of time in General Relativity to the natural, quantum mechanical clock of the dual theory, the modular clock, obtained by tracing out the ``observer''. It, thus, seems to enjoy a degree of universality that may extend its validity beyond the AdS/CFT context.

\section{Boundary description of bulk observers}
\label{sec:observers}
In this paper, we model a bulk observer as a black hole. It will be a large black hole, in the sense of having a Schwarzschild radius of order the AdS curvature scale, as required for having a simple associated modular flow. However, we will take it to be much smaller than the features of the spacetime that it is intended to probe. For this reason our construction captures AdS scale locality and geometric features. 

It is, of course, very interesting to generalize the approach to small black holes, whose associated modular flows are less universal. The microcanonical eternal black holes of \cite{Marolf:2018ldl} are the natural candidate for such a generalization, since they are semiclassical configurations black holes whose horizon radius can be parametrically smaller than $L_{AdS}$ while being thermodynamically dominant within an energy window. We discuss this generalization and related subtleties in some detail in Section \ref{sec:discussion}. 

This Section summarizes the advantages of the probe black hole description and introduces the concepts we will utilize in articulating our proposal in Sections \ref{sec:time} and \ref{sec:particles}: the modular Hamiltonian, modular Berry transport and the observer's code subspace degrees of freedom.

\begin{figure}[t]
\includegraphics[width=16cm]{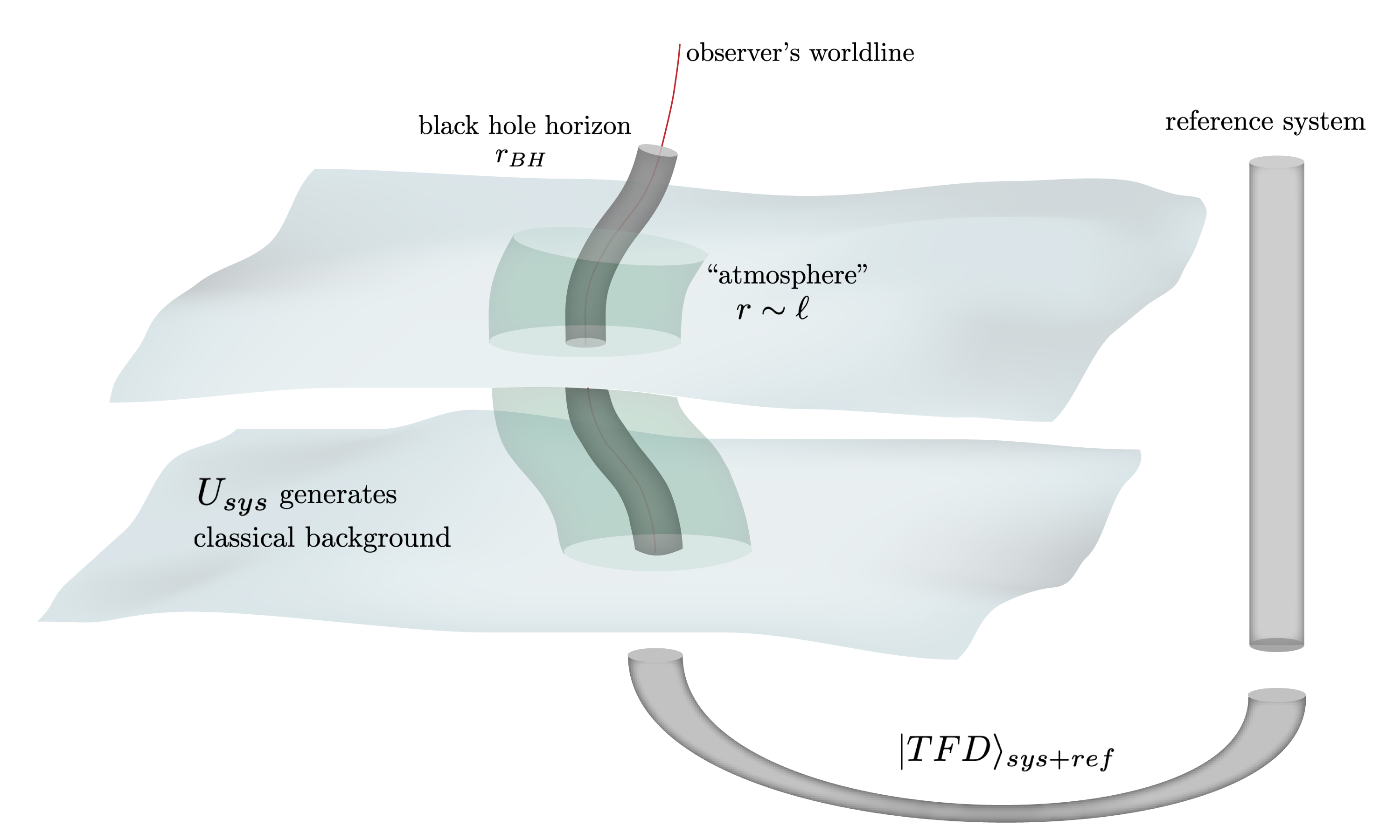}
\caption{\footnotesize{An illustration of our setup. The red line represents the worldline of an ideal observer. We replace them by a small black hole of radius $r_{BH}$ much smaller than the spacetime's curvature features which we thermally entangle with a reference system, assumed to be another AdS black hole for simplicity. The black hole propagates along the original geodesic due to the equivalence principle. The green tube of radius $\ell$ around the black hole represents its ``atmosphere'': The operators our observer can manipulate at any given time.}}
\label{setup}
\end{figure}

\subsection{A black hole ``observer''}
\label{subsec:BH}
The textbook observer in General Relativity is a local probe: They have zero size, no gravitational field and they travel along a worldline whose local neighborhood is approximately flat. This is a convenient idealization in the classical theory which, however, is unavailable quantum mechanically. Quantum observers are physical systems that should be included in the wavefunction of the Universe. They have finite mass and occupy volume of some linear size $\ell$ necessarily larger than their own Schwarzschild radius but smaller than the curvature scale of the spacetime they live in. Their gravitational field changes the local geometry around them to approximately Schwarzschild in their local inertial frame and quantum mechanical evolution entangles them with their environment ---a potential decoherence they need to protect themselves against to stay alive. Last but not least, observers have agency: They can manipulate and measure  degrees of freedom in their vicinity to perform experiments.

A general treatment of internal observers in AdS/CFT is a hopeless task but, fortunately, an unnecessary one. All we need is a more appropriate idealization of the physical observer above. In this paper, we squeeze our observer down to their Schwarzschild radius, collapsing them to a black hole of size $r_{BH}$ which we thermally entangle with an external reference system (fig. \ref{setup}). This black hole ``observer'' will approximately propagate along the same worldline, leaving the physics at scales much larger than $r_{BH}$ unaffected. The only difference is that excitations that originally intercepted the worldline are now absorbed by our black hole: They get detected by our observer's device!  

The entangled external system provides a precise meaning to the reference frame of the observer. In particular, one could consider gravitational operators probing the system spacetime whose diffeomorphism dressing is attached to the reference boundary through the Einstein-Rosen bridge. These must be complicated operators acting on both system and reference quantum systems, whose action in a (non-factorized) code subspace approximates that bulk description. 

In this work, we will be able to phrase our final results without needing such operators, which agrees with the fact that the proper time along the probe's trajectory between known near-boundary initial and final configurations is diffeomorphism invariant without requiring the observer's frame. As such, only system framed atmosphere operators and the probe black hole's density matrix appear in formula for the proper time. However, it is interesting to compare the result to the clock of the reference system, which gives a simple way to understand our results. 

Assuming the reference is a copy of the original CFT for simplicity, and that $r_{BH}\sim L_{AdS}$, the initial global state describing a black hole inserted somewhere in a classical, asymptotically AdS background has the general form:
\myeq{|\Psi\rangle = {\cal Z}^{-1/2} \sum_n \,e^{-\beta E_n/2}\,  U_{sys}|E_n\rangle_{sys} |E_n\rangle_{ref}\label{genstate}}
where $U_{sys}$ is a unitary transformation that excites the bulk fields and metric to create the background our black hole will propagate in.\footnote{Black holes with $r_{BH}\ll L_{AdS}$ could also in principle be discussed in our formalism, with the difference that the system-reference state would be more complicated since small black holes do not dominate the canonical ensemble. See Section \ref{sec:discussion} for related discussion. } Note that any state which looks like the eternal black hole in the  {\it reference}  system must be of this form. 

Crucially, collapsing our observer to a black hole offers protection from decoherence: Its degrees of freedom will remain spatially localized in the bulk for a time-scale of order the evaporation time which we will assume to be much larger than the time-scales of the experiments we will perform. 

We choose a black hole ``atmosphere'' of size $\ell$ to be the observer's lab: Local operators in the atmosphere can be directly manipulated or measured. We will assume $r_{BH}\lesssim L_{AdS} \ll \ell \ll L$ where $L$ the scale of the spacetime curvature perturbations about AdS. The CFT duals of the atmosphere operators at different times select a time-dependent subsystem of the boundary theory. This family of (abstract) CFT subsystems serves as a quantum notion of our observer's local frame which will be important in our formalism and is explained further in Section \ref{subsec:codesub}.

\subsection{Tracing out the observer: Modular Hamiltonian and Berry transport} 
\label{subsec:KandW}

An observer's description of the Universe they inhabit does not include a description of themselves. The degrees of freedom that make up the observer, which are generally entangled with the rest of the bulk, ought to be traced out. For the idealized eternal black hole observer of Section \ref{subsec:BH} this corresponds to tracing out the reference system.

\paragraph{Modular Hamiltonian}
Entanglement leads to uncertainty about a subsystem's quantum state. In our setup, ignorance of the state of the reference results in a mixed state $\rho=\text{Tr}_{ref}\left[|\Psi\rangle \langle \Psi|\right]$ for our black hole system. The Hermitian operator $K= -\log \rho$ is the \emph{modular Hamiltonian} and it defines an automorphism of the operator algebra. The action of this automorphism is generally non-local, except in special situations, and it can be thought of as generating the ``time'' evolution with respect to which the subsystem is in equilibrium. Somewhat more formally \cite{Witten:2018zxz}, the modular Hamiltonian of a quantum field theory subsystem in the state $\Psi$ is defined as $K=-\log \Delta_{\Psi}$ where $\Delta_\Psi$ is the ``KMS operator'' of $\Psi$ satisfying:
\myeq{\langle \Psi| \phi_1^\dagger\, \Delta_\Psi \,\phi_2 |\Psi\rangle = \langle \Psi| \phi_2  \,\phi_1^\dagger |\Psi\rangle \label{modKMS}}
for all correlation functions of the subsystems's operator algebra.

For our class of states (\ref{genstate}), the modular Hamiltonian is unitarily equivalent to the dynamical Hamiltonian of the boundary theory:
\myeq{K=2\pi \, U_{sys} H U_{sys}^{\dagger}}
where we have renormalized $H$ to $\frac{\beta}{2\pi}H$, for convenience. Recall that this is actually the most general type of state in which the full spacetime is identical to empty AdS-Schwarzschild on the {\it reference} side. In Schrodinger picture, time evolution acts on the state, resulting in a time-dependent modular Hamiltonian $K(t)$. 

A simple example we will frequently use to illustrate the ideas of this paper is a boosted black hole, which bounces back and forth in AdS. This is prepared by acting on a static black hole with the asymptotic boost symmetry of the spacetime, generated by the conformal boost $B$ in the CFT. Its dual state corresponds to (\ref{genstate}) with the choice $U_{sys}= e^{-iB\eta}$, where $\eta$ the black hole's rapidity, and the time dependent modular Hamiltonian reads explicitly:
\myeq{(2\pi)^{-1} K(t)= \cosh{\eta}\, H + \sinh\eta (P\cos t +B \sin t) \label{KboostedBH}}
where $H,P,B$ the conformal generators satisfying the usual $SL(2, \mathbb{R})$ algebra:
\myeq{ [B,H]=i P\, , \,\,\,\,\, [B,P]= i H \, ,\,\,\,\,\, [H,P]=iB}

\paragraph{Modular Berry Wilson lines}

A continuous family of modular Hamiltonians $K(t)$ ---like eq. (\ref{KboostedBH})---  selects a continuous family of bases in the Hilbert space, consisting of the eigenvectors of $K(t)$. The local generator $D(t)$ of the basis rotation can generally be obtained as the solution to the problem:
\myal{\partial_t K(t)&= -i[D(t), K(t)] \label{transport1}}
Equation (\ref{transport1}) by itself determines $D(t)$ only up to modular zero modes $Q(t)$, generating symmetries of the reduced state $[Q(t),K(t)]=0$. This reflects the freedom to choose at will the  local modular frame, e.g. phases of eigenstates of different $K(t)$, which is an example of a Berry phase, discussed in detail in \cite{Czech:2019vih}. A canonical map between the bases that is intrinsic to the family $K(t)$ can be constructed following Berry's footsteps \cite{Berry:1984jv}, by defining the \emph{modular parallel transport operator} as the solution to the problem (\ref{transport1}) supplemented by the condition:
\myal{P_0^t[D(t)]&=0\label{transport2}}
where $P_0^t$ is the projection of the Hermitian operator $D(t)$ onto the subspace of zero modes of $K(t)$. Eigenframes of different $K(t)$ are related by modular Wilson lines, the path ordered exponential of the parallel transport $D(t)$ 
\myeq{{\cal W}(t_1,t_2) = {\cal T} \exp \left[ -i \int_{t_1}^{t_2} dt'\, D(t')\right] \label{modW}}
$\cal{W}$ can also be thought of as a canonical unitary automorphism of the operator algebra of our system:
\myeq{ O_{\cal W}(t) =  {\cal W}(0,t) \, O \, {\cal W}^\dagger (0,t) \label{Wautom}}
with the properties:
\myal{ K(t)&= {\cal W}(0,t) \, K(0) \, {\cal W}^\dagger (0,t)  \label{Wautomprop1} \\
\langle \Psi(t)| \,O^{(1)}_{\cal W}(t)&\dots O^{(n)}_{\cal W}(t) \, |\Psi(t)\rangle = \langle \Psi | \,O^{(1)} \dots O^{(n)}\, |\Psi\rangle \label{Wautomprop2}}

When $K(t)$ is obtained via the time evolution of the system ---as in the example of the previous Subsection--- the modular parallel transport is related to the, possibly time-dependent, Hamiltonian $H(t)$ via:
\myeq{D(t)= H(t) - P_0^t[H(t)] \label{DforH}}
In our boosted black hole example with modular Hamiltonian (\ref{KboostedBH}), the parallel transport problem can be solved explicitly by a straightforward group theory exercise. A convenient way to express the solution is:
\myal{ D(t)& = \dot{x}(t) \, P + \dot{\eta}(t) e^{-iP\,x(t)} B \,e^{iP\,x(t)} - b(t) \,K(t) \label{boostedD}\\
x(t)&= \tanh^{-1} \left( \tanh \eta(0) \, \sin t \right) \label{bhposition}\\
\eta(t)&= \sinh^{-1} \left( \sinh \eta(0) \cos t\right) \label{bhrapidity}\\
b(t)&= \frac{1}{2\pi} \dot{x}(t) \sinh \eta(t) \label{bhzeromode}}
where $x(t), \eta(t)$ measure the black hole's geodesic distance from the AdS origin and its rapidity in the global frame respectively, while the last term enforces the vanishing of the zero-mode component of $D(t)$. The modular parallel transport operator (\ref{modW}) then reads:
\myeq{{\cal W}_{\text{boosted BH}}(0,t) = {\cal T} \exp\left[ -i\int_0^t dt'\, D(t') \right] = e^{-iPx(t)} e^{-iB\left(\eta(t)-\eta(0)\right)} e^{\frac{i}{2\pi}K(0) \,\int_0^t dt'\,\dot{x} \sinh \eta(t')}\label{Wboosted}}

\paragraph{Modular holonomies.} A key property of the modular Berry transport is that it generally leads to non-trivial holonomies ---a fact that will play an important role in our subsequent discussion. We can consider two families of modular Hamiltonians, $K_1(t), K_2(t)$ for $t\in [0,T]$, which coincide at the initial and final times: $K_1(0)=K_2(0)$ and $K_1(T)=K_2(T)$. These could correspond to two distinct worldlines for our black hole that begin and end at the same spacetime location and with the same momentum. $K_1(t), K_2(t)$ then form a closed ``loop'' 
\myeq{ K(t) = \begin{cases} K_1(t) &0\leq t\leq T  \\ K_2(2T-t), &T\leq t \leq2T   \end{cases}}
and the property (\ref{Wautomprop1}) becomes:
\myeq{K(0)={ \cal W}_{loop}(0,2T) \,K(0)\,{ \cal W}_{loop}^\dagger (0,2T) }
which implies that the modular Wilson loop, ${\cal W}_{loop}(0,T)$ will be a, generally non-trivial, element of the modular symmetry group, generated by the zero modes $Q_i(0)$:
\myeq{{\cal W}_{loop}(0,T)= \exp\left[-i \sum_i  d_i\, Q_i(0)\right] \label{MBloop}}
This is a modular Berry holonomy, an example of which we will see below in our discussion of time dilation between observers.

\subsection{The observer's code subspace}
\label{subsec:codesub}
Up to this point, we have treated the bulk observer as a physical system, entangled with their environment in the global wavefunction, which results in a time-dependent modular Hamiltonian $K(t)$ upon tracing them out. Another defining characteristic of an observer, however, is their ability to control  some degrees of freedom in their Universe to learn about Its state. 
In our model, these will be the local bulk fields in a small atmosphere of size $\ell$ around the black hole, denoted by $\phi_i$ with $i$ an abstract index, and $O(1)-$degree polynomials built out of them and their derivatives. 

The atmosphere degrees of freedom on a particular bulk timeslice $\Sigma_t$ which asymptotes to boundary time $t$ form a set of observables 
\myeq{S_t\equiv \{ \phi_i(x_i), \phi_j(x_j)\phi_l(x_l), \partial \phi_k(x_k), \dots \big| x_i \in \Sigma_t \,\, \text{and} \,\, \, |x_i -x_H|\leq \ell  \}}
By acting with elements of $S_t$ on the background state (\ref{genstate}) we obtain the observer's instantaneous code subspace \cite{Almheiri:2014lwa}: The subspace of the CFT Hilbert space the observer can explore with their apparatus at a given bulk time. Crucially, this code subspace is not generally preserved by time evolution. The observer moves in the bulk hence the operators in their vicinity (and their boundary duals) differ at different times, resulting in an evolution of the CFT subspace that the observer can probe.

As remarked in our introductory Section, we will assume knowledge of the CFT duals of the atmosphere operators at an initial and a final timeslice, $S_{t_i}$ and $S_{t_f}$ respectively ---but not in-between. This is physically reasonable when studying processes where the black hole is introduced far out in the asymptotically AdS region and returns to it at some later boundary time, in which case the familiar HKLL prescription for an AdS black hole can be employed for the initial and final reconstruction.

\paragraph{Dressing} $S_{t_i}$, $S_{t_f}$ refer to local operators in a theory of gravity so it is important to clarify their gravitational framing. The choice of an initial and final timeslices $\Sigma_{t_i}$, $\Sigma_{t_f}$ in the definition of the operator sets is a selection of a bulk gauge, at least in the vicinity of the black hole. Since the black hole is assumed to be near the AdS boundary at those moments, its local neighborhood is diffeomorphic to an AdS-Schwarzschild geometry. These local AdS-Schwarzschild coordinates serve as the analog of the local inertial frame about an idealized observer's worldline. We are interested in describing the operators in the black hole's reference frame, thus we choose $\Sigma_{t_i}$, $\Sigma_{t_f}$ to both be constant time with respect to the corresponding local time-like killing vector within the atmosphere (fig. \ref{BHpropertime}). Operators $\phi \in S_{t_i}$ or $S_{t_f}$ can then be labelled by their location in this local AdS-Schwarzschild coordinate system. 

These are operators that are dressed with respect to the AdS boundary with the property that their action within the code subspace results in their insertion at given positions relative to the ``local horizon'', namely the place where the horizon would form if no matter were absorbed in the future. This ``local horizon'' may generally differ for $S_{t_i}$ and $S_{t_f}$, as we will see in Section \ref{sec:particles}. Using standard HKLL, we can construct such  operators that work in an entire family of perturbative excitations about a black hole of a given temperature \cite{Hamilton:2006fh, Kabat:2013wga}. 

It is important to note that when acting within the code subspace of small perturbations\footnote{The perturbations must be small at the specified time, to avoid exciting scrambling modes that lead to large deviations ---as we explain in Section \ref{sec:particles}.} around a given semi-classical spacetime state, bulk operators with different dressings that result in insertion at the same point in the original spacetime are equal at leading order. The associated states $\phi |\Psi\rangle$ would appear to have different gravitational field configurations associated to the energy of the particle produced by $\phi$, however these are subleading to the quantum fluctuations in the ambient gravitational field. This can be seen by explicitly computing the overlap of two such states with different dressings for $\phi$ that classically result in the same insertion point. The states are identical as quantum states up to $G_N$ corrections. 

\paragraph{Example} For illustration, we return to our boosted black hole example (\ref{KboostedBH}). The atmosphere operators at the $t=0$ global AdS timeslice, when the black hole is located at the AdS origin and has rapidity $\eta$, can be obtained from the standard HKLL operators in a static AdS-Schwarzschild metric via the action of a boundary  conformal boost:
\myeq{ \phi^0(r,\Omega) = e^{-iB\eta} \phi^{\text{static}} (r,\Omega) \,e^{iB\eta} \,\,\text{where: } \ell_{Pl}\ll r-r_{BH}<\ell \label{boostedphi0}}
After global time $t$ the black hole has moved to a new location $x(t)$ and has a local rapidity $\eta(t)$ given in (\ref{bhposition}) and (\ref{bhrapidity}). By the previous reasoning, the atmosphere observables, \emph{in the Schrodinger picture}, are given by:
\myeq{\phi^{t}(r,\Omega) =e^{-iPx(t)} e^{-iB\eta(t)} \phi^{\text{static}} (r,\Omega) \,e^{iB\eta(t)}e^{iPx(t)} \,\,\text{where: } \ell_{Pl}\ll r-r_{BH}<\ell \label{boostedphit}}

\paragraph{Proper time evolution} The unitary that relate the atmosphere operators $S_t$ at different times $t$ is, by definition, the proper time evolution along the black hole's worldline. We will denote this unitary by $V_S(t_1, t_2)$ or $V_H(t_1, t_2)$ depending on whether we represent it in the Schrodinger or Heisenberg picture. The goal of the remainder of this paper is to understand the construction of $V$ directly in the CFT language, without reference to bulk reconstruction ---except at the initial and final moments of our probe black hole's history.

\section{The holographic measurement of time}
\label{sec:time}
With all the necessary concepts in place, we are ready to present the advertised connection between modular time and proper time, when no matter gets absorbed by our probe black hole; the case of particle absorption is postponed for Section \ref{sec:particles}.

\subsection{The proposal: Proper time from modular time}
\label{subsec:proposal}

We now provide three complementary perspectives on our main claim. We start with a bulk geometric argument that offers some useful intuition and then make the case quantum mechanically, using both Heisenberg and Schrodinger picture reasoning, each of which illuminates different aspects of the physics.

\begin{figure}
\begin{center}
\includegraphics[width=15cm]{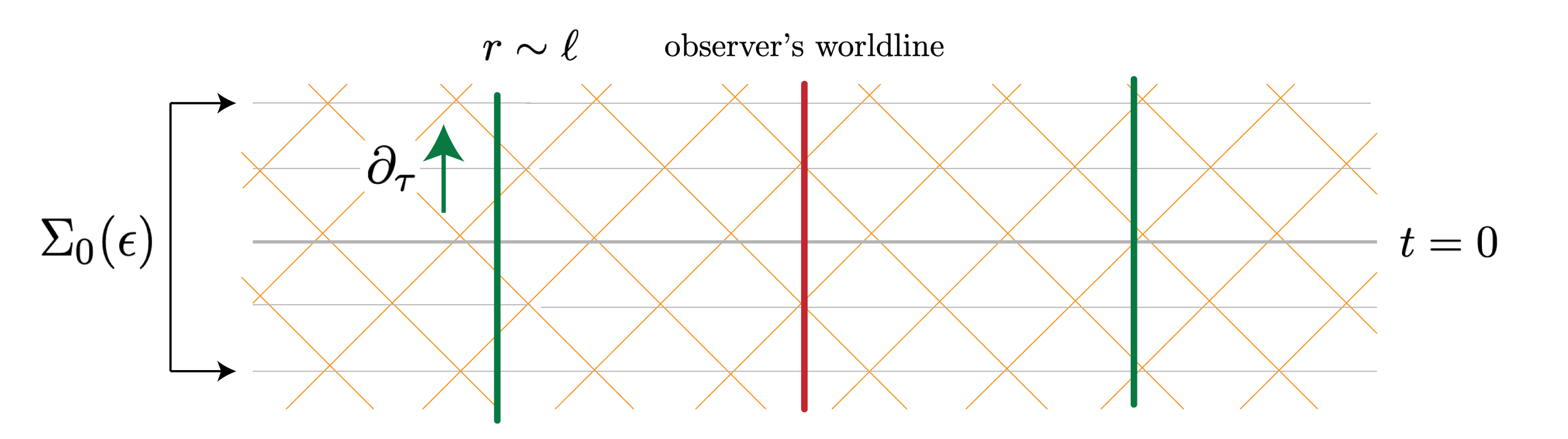}
\includegraphics[width=15cm]{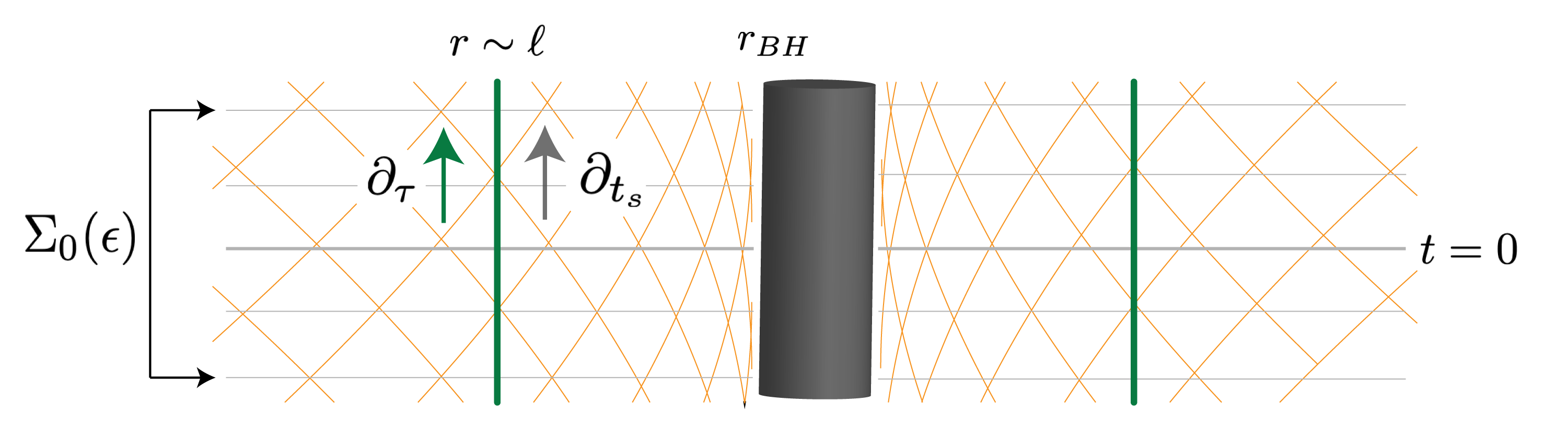}
\caption{\footnotesize{Our black hole is introduced geometrically by cutting a hole of size $\ell$ around the ideal observer's worldline in the initial Cauchy slice and a small time band $\Sigma_0(\epsilon)$ about it and replacing the interior with a black hole metric. The local killing vector generating the worldline's proper time is glued to the local generator of Schwarzschild time which, in turn, is modular time. As long as nothing falls in the black hole, this identification is valid everywhere along the worldline, suggesting that modular time is correlated to proper time.}}
\label{BHpropertime}
\end{center}
\end{figure}

\paragraph{An intuitive geometric argument} In the bulk, our black hole observer can be understood via the geometric construction of fig \ref{BHpropertime}. We start with an idealized probe observer of some small mass $m$ and we choose an initial Cauchy slice $\Sigma_0$ which is ``constant time'' in their local inertial frame, as well as the Cauchy slices within an $\epsilon\to 0$ thickness time band $\Sigma_0(\epsilon)$ around it. The geometry of this time band near the observer's location reads, in local inertial coordinates:
\myeq{ds_{obs}^2\,\overset{r\gg \mu}{\approx}\, -d\tau^2 +dr^2  + r^2 d\Omega + \frac{\mu}{r^{d-3}} (d\tau^2 +dr^2) +O\left((Lx)^2\right)}
where $\tau$ is the proper length of the worldline, $r$ the radial distance from it and $L$ is the scale of the curvature features of the surrounding spacetime. 

We cut a hole of size $\ell$ in $\Sigma_0(\epsilon)$, with $\mu^{\frac{1}{d-3}}\ll \ell \ll L$, around the worldline and replace its interior with the black hole geometry \footnote{The geometry in the atmosphere of the $L_{AdS}$ sized black holes described the thermofield double state (\ref{genstate}) is, instead, given the AdS-Schwarzschild metric which needs to be glued to the local AdS frame of the observer's neighborhood. This subtlety has no effect on the argument of this section and is only omitted for clarity.}:
\myeq{ds_{BH}^2 \,\overset{r_s<\ell}{=}- \left(1-\frac{\mu}{r^{d-3}_s}\right)dt_s^2 + \left(1-\frac{\mu}{r^{d-3}_s}\right)^{-1} dr_s^2 +r_s^2 d\Omega}
To ensure a smooth gluing at $\ell$, the radial coordinates must be identified $r=r_s$ since they control the size of the transverse sphere at $r=r_s=\ell$. Similarly, the local Schwarzschild killing vector $\xi_s=\partial_{t_{s}}$ at $r=\ell- \delta$ gets identified with the corresponding local inertial frame killing vector $\xi_\tau=\partial_{\tau}$ at $r=\ell +\delta $ for $\delta \to 0$. We then feed these initial conditions to the Einstein equations and evolve the system. 

By the equivalence principle, this black hole will propagate along the original worldline, as long as the scale of bulk curvature features is much larger than $r_{BH}$. Due to our assumption of no infalling energy, its atmosphere will also remain locally diffeomorphic to Schwarzschild, so at any point along its path, and at distances $r_{BH}\ll r \sim \ell \ll L$ the Schwrarzschild clock $t_S$ will coincide with the local inertial clock $\tau$ of the idealized observer. 

To complete the argument, we need to relate the Schwarzschild clock to modular time. The CFT modular Hamiltonian for a global state $|\Psi\rangle$ is identified with the bulk modular Hamiltonian \cite{Jafferis:2014lza, Jafferis:2015del} which, in turn, is defined via the KMS operator (\ref{modKMS})
\myeq{\langle \Psi| \phi_1^\dagger e^{-K} \,\phi_2 |\Psi \rangle = \langle \Psi| \phi_2 \,\phi_1^\dagger|\Psi\rangle }
As long as particles do not cross paths with our black hole's trajectory, the state of the black hole atmosphere will remain in an approximate local thermal equilibrium: Expectation values of atmosphere observables will be approximately given by their thermal ones, with the Wick rotation of the local timelike killing direction $t_s$ playing the role of the thermal circle. By virtue of the usual KMS condition then, we have
\myeq{ \langle \Psi| \phi_1^\dagger e^{-K} \,\phi_2 |\Psi \rangle = \langle \Psi| \phi_2 \,\phi_1^\dagger|\Psi\rangle \approx  \langle \Psi| \phi_1^\dagger e^{- 2\pi P_{\xi_s}} \,\phi_2 |\Psi \rangle}
where: $P_{\xi_s}$ the geometric generator of the geometric flow of $\xi_s$. Hence, within the local thermal atmosphere $r\lesssim \ell$, $K$ acts like the geometric generator $2\pi P_{\xi_s}$ which coincides with the worldline proper time generator $2\pi P_{\xi_\tau}$ at $r\sim \ell$.

\paragraph{Heisenberg picture} To justify our proposal quantum mechanically, it is simplest to work in the Heisenberg picture. Hamiltonian evolution of the system is described by the unitary rotation of the operator basis, while the state and by extension the modular Hamiltonian remain fixed. The atmosphere operator set $S_t$ of Section \ref{subsec:codesub}, however, does not simply consist of the Heisenberg evolved elements of $S_0$, because their correct evolution, which we denote by the unitary $V_H(0, t)$, needs to also reflect the motion of the black hole in the gravity dual:
\myeq{\phi_H^t(x) = V_H(0,t) \,\phi_H^0(x) \,V_H^\dagger(0,t) \label{atmosphereisomorphism}}
where the subscript $H$ is introduced to make the Heisenberg picture explicit. Given our assumption that
\begin{enumerate}
\item the black hole on the initial and final timeslices $\Sigma_{0}, \Sigma_{t}$ is located in the asymptotic AdS region and is thus locally diffeomorphic to AdS-Schwarzschild, with the state being approximately invariant under the local killing time-like vector
\item no energy is absorbed by our probe black hole ---an assumption we lift in Section \ref{sec:particles}
\end{enumerate}
we conclude that correlation functions of Heisenberg operators in $S_0$ and $S_t$ are identical in the background state $|\Psi\rangle$:
\myeq{\langle \Psi|  \,\phi_{H,1}^t\dots \phi_{H,n}^t \, |\Psi \rangle = \langle \Psi | \,\phi_{H,1}^0 \dots \phi_{H,n}^0\, |\Psi\rangle \label{initialequalsfinalH}}

The meaning of these operators are bulk fields, dressed to the AdS boundary in such a way that in the subspace under consideration they are inserted in the atmosphere as labeled by coordinates relative to the extremal surface. Due to the Schwarzschild time-like isometry of the near horizon region, one needs to additionally specify a timeslice, anchored to the AdS boundary. We can do this because we assume that the black hole begins and ends its journey in understood regions near the boundary.

Due to (\ref{initialequalsfinalH}), the isomorphism  $V_H(0, t)$ must be a ``modular symmetry'', when acting on the observer's code subspace. Such a unitary can be generated by two classes of operators:
\begin{itemize}
 \item zero modes $Q_a$ of the modular Hamiltonian \emph{projected onto the observer's code subspace}:
\myeq{ [Q_a, P_{code} K(0) P_{code}]=0, \,\,\text{where: } {\cal H}_{code}= \{ O |\Psi\rangle, \, \forall \,\,O\in S_0\} \label{codezeromodes}}
\item operators $G_\lambda = G^\dagger_\lambda$ that are eigenoperators of the code subspace $K$ with imaginary eigenvalues
\myeq{[P_{code} K(0)P_{code}, G_\lambda]= -i\lambda G_\lambda \label{imaginaryG}}
\end{itemize}
The latter necessarily annihilate state $|\Psi\rangle$ since otherwise $G_\lambda |\Psi\rangle$ would constitute an eigenstate of the modular Hamiltonian with imaginary eigenvalue which contradicts the Hermiticity of $K$. A special class of these imaginary eigenvalue operators is those with $\lambda=\pm 2\pi$. These were dubbed modular scrambling modes in \cite{deBoer:2019uem} because they saturate the bound on modular chaos and they were argued to generate null translations near the entangling surface. The simplest example of such a scrambling mode is the Averaged Null Energy operator $\int dx^+ \,T_{++}( \Omega)$ at the horizon of a static AdS black hole in equilibrium, where the eigenvalue $2\pi i$ follows from the near horizon Poincare algebra.

We claim that $G_\lambda$ do not contribute to the unitary $V_H$ when no particles get absorbed by our black hole. This is not true for cases with non-vanishing infalling energy flux which, as we show in Section \ref{subsec:Kwithparticles}, results in a scrambling mode $G_{2\pi}$ contribution. Modes with $|\lambda|>2\pi$ are forbidden by the modular chaos bound \cite{deBoer:2019uem, Faulkner:2018faa}, as we review in Section \ref{subsec:modchaos}. We are unaware of any situations where $G_\lambda$ with $-2\pi < \lambda < 2\pi$ appear, thus we tentatively suggest they are, also, absent in general ---leaving a more thorough investigation of this issue for future work. With some foresight, we can return to the case with no absorption and express the evolution operator in (\ref{atmosphereisomorphism}) as:
\myeq{V_H(0,t) = \exp\left[ -i  \frac{\tau(t)}{2\pi}\, K(0) -i \sum_a  d_a(t) Q'_a \right] \label{VHfinal}}
where we separated the modular Hamiltonian from the rest of the zero modes $Q'_a$. \emph{We propose that the coefficient of the modular Hamiltonian $\tau(t)$ measures the proper time along the bulk observer's worldline, in units of the black hole temperature $\beta/2\pi$.} The other zero modes $Q_a'$ describe the precession of the symmetry frame of the observer, e.g. a certain amount of rotation of the local reference frame.

The intuition for identifying $\tau(t)$ with proper time is as follows. Within the code subspace, the action of the atmosphere $\phi$ is, at leading order, identical to bulk operators that are framed to the reference boundary, at an appropriate time. Evolution under the reference Hamiltonian moves the anchor point of those operators, and this gives the local Schwarzschild evolution in the atmosphere region. Thus the proper time along the trajectory is exactly the amount of modular evolution required to relate the initial and final atmosphere operators, where we equate operators with equal projection onto the code subspace.

\begin{figure}
\begin{center}
\includegraphics[width=13cm]{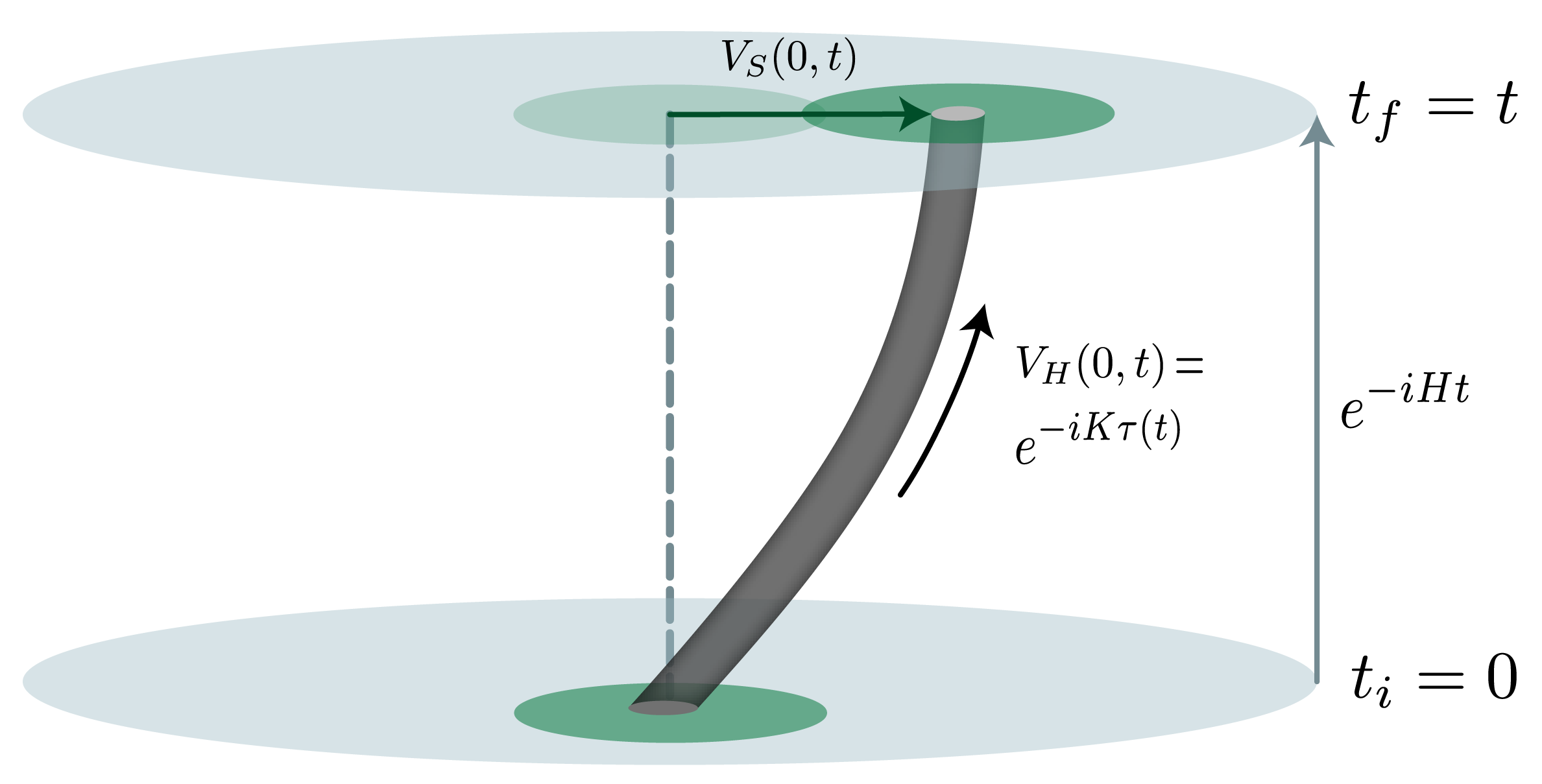}
\caption{\footnotesize{Illustration of the three different flows appearing in our discussion. $H$ is the CFT Hamiltonian generating global AdS evolution. $V_H$ is modular flow which maps the $t_i=0$ atmosphere operators (green disk on $t_i=0$ slice) to the Heisenberg picture atmosphere operators at $t_f=t$. $V_S$ describes the evolution of the atmosphere operators in the Schrodinger picture and captures the motion of the black hole relative to the boundary.}}
\label{evolutionillustration}
\end{center}
\end{figure}

\paragraph{Schrodinger picture} It is illuminating to present the same argument in the Schrodinger picture,  where  the black hole state in what evolves under the Hamiltonian evolution, as encapsulated in a time-dependent $K(t)$. While, now, the operator basis does not evolve, the atmosphere operator set $S_t$ does, due to the motion of the bulk black hole relative to the AdS boundary. The Schrodinger picture atmosphere operators in $S_0$ and $S_t$ are related by a unitary $V_S(0, t)$:
\myeq{\phi^t(x) = V_S(0,t) \,\phi^0(x) \,V_S^\dagger(0,t) \label{atmosphereisomorphism}}
where the subscript $S$ is a reminder that we are working in the Schrodinger picture. 

The Schrodinger version of eq. (\ref{initialequalsfinalH}) is that correlation functions of operators in $S_0$ in the initial state $|\Psi\rangle$ are equal to correlation functions of the final atmosphere operators $S_t$ in $|\Psi(t)\rangle$:
\myeq{\langle \Psi (t)|  \,\phi^t_1\dots \phi^t_n \, |\Psi(t)\rangle = \langle \Psi | \,\phi^0_1 \dots \phi^0_n\, |\Psi\rangle \label{initialequalsfinal}}

By virtue of eq. (\ref{Wautomprop2}), property (\ref{initialequalsfinal}) the isomorphism (\ref{atmosphereisomorphism}) can be identified with the modular Berry transport ${\cal W}$, up to a symmetry $Z_Q$ of the observer's code subspace correlators:
\myal{V_S(0,t)&= {\cal W}(0,t)\, Z_Q[c_a(t)] \label{VSexp} \\
\text{where: } Z_Q[c_a(t)] &=\exp \left[ -i \sum_a c_a(t) Q_a(0) \right] \label{Zexpression}}
and, as before, $Q_a$ are the code subspace modular zero-modes (\ref{codezeromodes}). 

As explained in Section \ref{subsec:KandW}, for a family $K(t)$ obtained by Hamiltonian time evolution ${\cal W}$ is generated by (\ref{DforH}) so (\ref{VSexp}) becomes:
\myal{V_S(0,t) &= {\cal T} \exp\left[-i \int_0^t \,dt' \,(H - P_0^{t'}[H])\right] Z_Q\left[c_a(t)\right]\nonumber\\
&= e^{-iHt} \exp\Big[ i\,\int_0^t dt'\, e^{iHt'} P_0^{t'}[H]e^{-iHt'} \Big] Z_Q\left[c_a(t)\right] \label{VSexpfinal}} 
Substituting (\ref{VSexpfinal}) in eq. (\ref{atmosphereisomorphism}) and \emph{switching to the Heisenberg picture} we get the following relation between the atmosphere operators:
\myal{\phi^t_H (x) &= V_H(0,t) \,\phi_H^0(x) \, V^\dagger_H(0,t)\nonumber\\
\text{where: } V_H(0,t)&= \exp\Big[ i\,\int_0^t dt'\, e^{iHt'} P_0^{t'}[H]e^{-iHt'} \Big] Z_Q\left[c_a(t)\right]  \label{atmisomorphismH}}
The unitary $V_H$ is now obtained by the product of two contributions, one coming from the zero-mode projection of the CFT Hamiltonian and the other from the code subspace symmetry transformation $Z_Q$ in (\ref{VSexp}). These two terms have distinct physical interpretations which we discuss in the context of our AdS example below. This decomposition will be important in our discussion of the \emph{relative time} between two observers in Section \ref{subsec:timedilation}, where the $Z_Q$ contributions will give rise to a modular Berry holonomy, providing a conceptually clean way of organizing the CFT dual of time dilation.

\subsection{A test case: Moving black holes in AdS}
\label{subsec:adsBH}

As an illustration of the idea, we focus on black holes moving in empty AdS along \emph{arbitrary} worldlines and compute their proper time using our proposed method. 

\paragraph{AdS Black holes in inertial motion} Consider the case of the boosted black hole, propagating along an AdS geodesic. In the CFT, it is characterized in the Schrodinger picture by the time-dependent modular Hamiltonian (\ref{KboostedBH}), with atmosphere operators on the initial and final timeslices  given by (\ref{boostedphi0}) and (\ref{boostedphit}) respectively. The unitary $V_S(0,t)$ in eq. (\ref{atmosphereisomorphism}) is equal to:
\myeq{V_S(0,t) = e^{-iPx(t)} e^{-iB\left(\eta(t)-\eta(0)\right)} \label{VSboostedbh}}
Recalling the expression (\ref{Wboosted}) for the modular parallel transport in this example, $V_S$ can be written as:
\myeq{V_S(0,t) ={\cal W}_{\text{boosted BH}}(0,t) \exp \left[-i(2\pi)^{-1}K(0) \int_0^t dt'\, \dot{x}(t') \sinh \eta(t')\right] \label{VSboosted}}
Equally straightforwardly, we can compute the projection of the dynamical Hamiltonian on the modular zero modes of $K(t)$, which reads:
\myeq{ P_0^t[H]= \frac{1}{2\pi}\cosh \eta(0) K(t) \label{Hzeromodeboosted}}

Combining the results (\ref{VSboosted}) and (\ref{Hzeromodeboosted}) in expression (\ref{atmisomorphismH}) for the proper time evolution operator $V_H(0,t)$ we find:
\myeq{V_H(0,t) = \exp \left[ -i(2\pi)^{-1} K(0) \, \int_0^t dt'\,\left( \dot{x}(t') \sinh \eta(t') - \cosh \eta(0)\right)\right] \label{boostedfinalresult}}
The coefficient of the modular Hamiltonian, using the expressions (\ref{bhposition}) and (\ref{bhrapidity}) for $x(t)$ and $\eta(t)$, reads:
\myeq{ \tau(t) = \tan^{-1} \frac{\tan t}{\cosh \eta(0)} \label{propertimeboosted}}
which is indeed the proper length of the black hole's worldline between the $0$ and $t$ global AdS timeslices.

\paragraph{A worldline interpretation of the result}  At a sufficiently coarse-grained level, our black hole behaves like a particle, whose propagation in the bulk spacetime follows from extremization of its worldline action, i.e. its proper length
\myeq{S_{\text{worldline}}[x^\mu(t)] = \int d\tau = \int_0^t dt' {\cal L} (x^\mu(t),\dot{x}^\mu(t) | g)\label{worldline}}
which can alternatively be written as a Legendre transform of the worldline energy $E[x^\mu(t)]$:
\myeq{S_{\text{worldline}}[x^\mu(t)] = \int_0^t dt\,\left( \dot{x}^\mu  \frac{\delta {\cal L}}{\delta \dot{x}^\mu} - E[x^\mu(t)]\right)}

It is instructive to observe that the two zero mode contributions to $V_H$ in eq. (\ref{atmisomorphismH}) have different physical interpretations. The zero mode of the CFT Hamiltonian (\ref{Hzeromodeboosted}) measures the the energy of the black hole $E[x^\mu(t)]$, namely the worldline Hamiltonian evaluated on-shell, while the zero mode contribution to (\ref{VSboosted}) in the chosen gauge is equal to the quantity $\dot{x}^\mu(t) \frac{\delta {\cal L}}{\delta \dot{x}^\mu}$ along the trajectory. The two are combined in eq. (\ref{boostedfinalresult}) to give an amount of modular evolution equal to the on-shell worldline action for our probe black hole.

\paragraph{Accelerating AdS black holes} The example can be extended to arbitrary accelerating black holes. A simple example is a black hole that starts at the AdS origin at $t=0$ with rapidity $\eta(0)$ and at some boundary time $t_0$ receives a kick that changes its rapidity, e.g. flips it from $\eta(t_0)$ to $-\eta(t_0)$. The black hole returns to the origin at global time $t=2t_0$ when its internal clock is showing $\tau(2t_0) = 2\tan^{-1}\frac{\tan t_0}{\cosh \eta(0)}$, according to the bulk calculation. 

The modular Wilson line associated to the corresponding family of modular Hamiltonians can be computed straightforwardly from its defining equations (\ref{transport1}), (\ref{transport2}): 
\myal{{\cal W}(0,2t_0)& = {\cal T} e^{-i\int_0^{2t_0} dt'\, D(t') } \nonumber\\
&= {\cal W}_{\text{boosted BH}} (\pi-t_0,\pi) \exp\left[2iB_{x(t_0)} \eta(t_0)\right] {\cal W}_{\text{boosted BH}} (0,t_0) \label{accelWilson}}
where ${\cal W}_{\text{boosted BH}}$ is given by (\ref{Wboosted}), and the instantaneous boost $B_{x(t_0)}= e^{-iPx(t_0)} B\,e^{iPx(t_0)} $ accounts for the $t=t_0$ discontinuity in the operator family $K(t)$ due to the kick of the black hole. This discontinuity is, of course, an artifact of our approximation that would be absent from any realistic accelerating black hole. 

On the boundary, the local atmosphere fields at $t=0 $ and $t=2t_0$ are related by
\myal{\phi^{2t_0} &= e^{2 iB\eta(0)} \phi^0 e^{-2iB\eta(0)} \label{accelatmops} }
In view of (\ref{accelWilson}), the map $V_S(0,2t_0)=e^{2iB\eta(0)}$ in (\ref{accelatmops}) can be shown to be equal to
\myeq{V_S(0,2t_0)= {\cal W}(0,2t_0)\, \exp\left[-2i(2\pi)^{-1}K(0) \int_0^{t_0} dt'\, \dot{x}(t') \sinh \eta(t')\right]}

Extracting the proper time requires computing the Heisenberg picture evolution operator (\ref{atmisomorphismH}). The zero mode component of the CFT Hamiltonian is once again given by (\ref{Hzeromodeboosted}) so the final result reads
\myeq{V_H(0,2t_0)= \exp \left[-2i\tan^{-1} \frac{\tan t_0}{\cosh\eta(0)}\, \frac{K(0)}{2\pi}\right] }
which agrees with the bulk geometric computation. 

By an appropriate dense sequence of small kicks like the one studied here, an arbitrary worldline can be constructed, allowing our method to correctly compute the proper length of any timelike path in AdS. This construction guarantees that our prescription works in all weak curvature perturbations of Anti-de Sitter spacetime.

\begin{figure}
\begin{center}
\includegraphics[width=7.5cm]{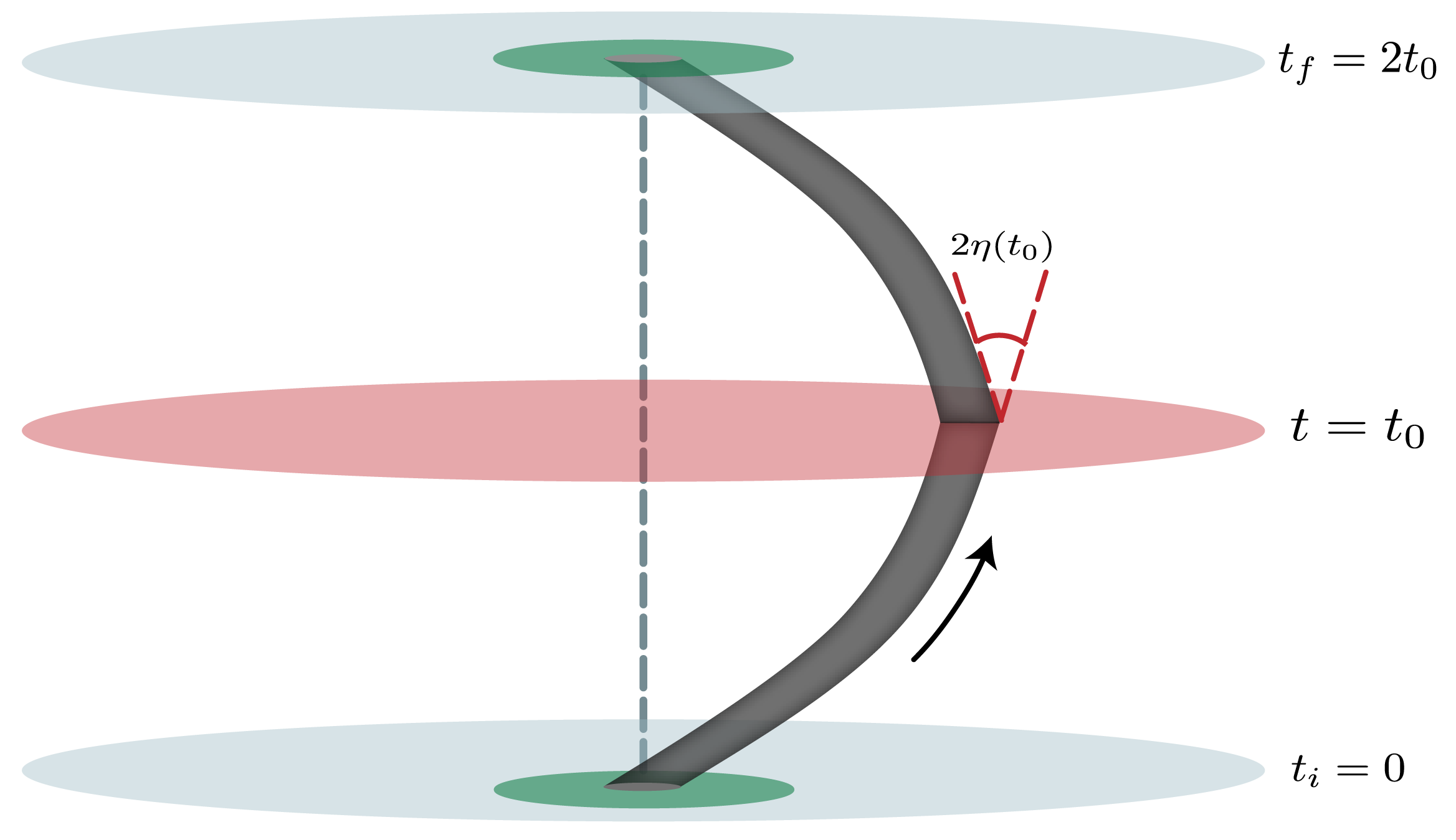}
\includegraphics[width=7.5cm]{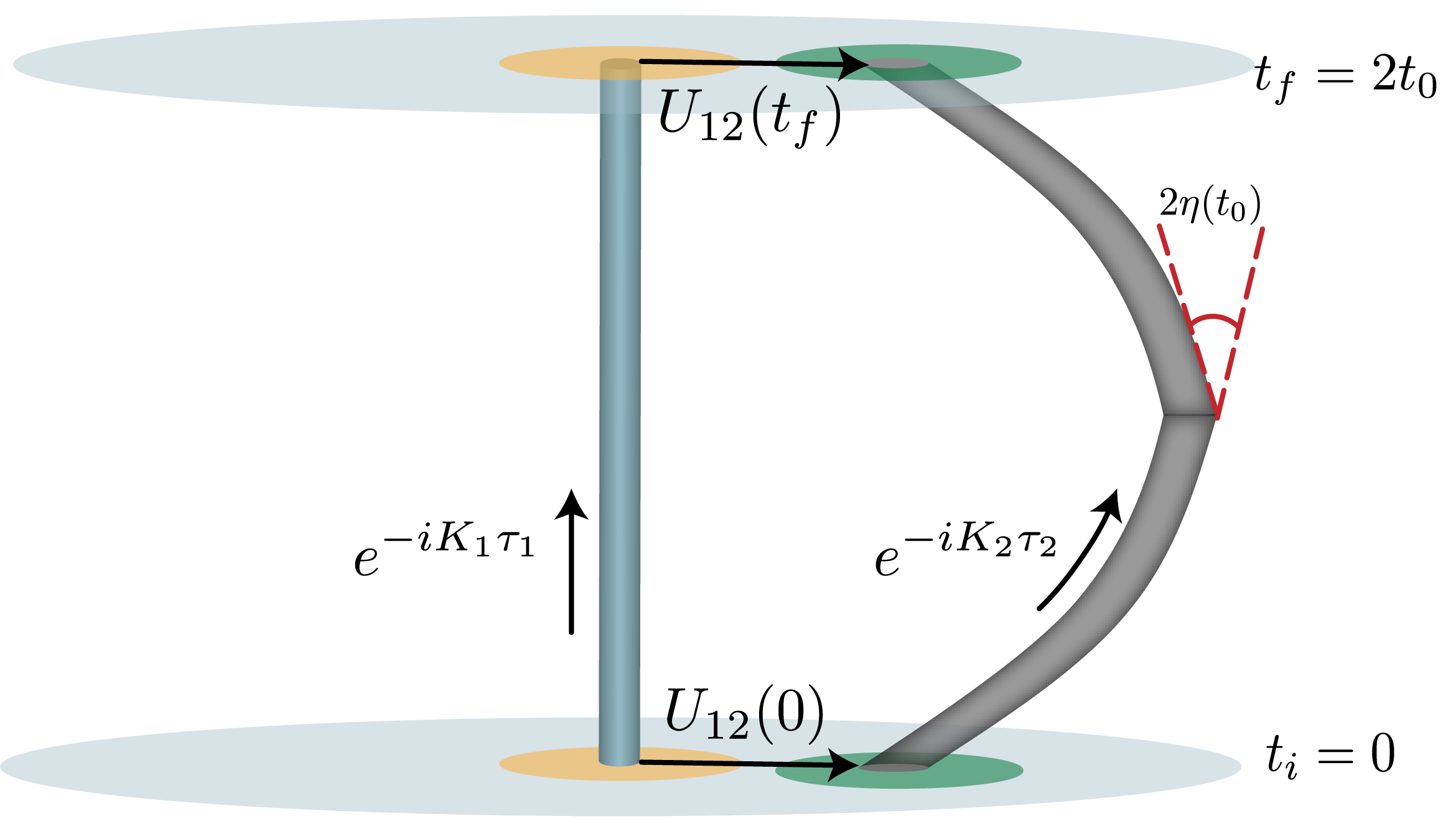}
\caption{\footnotesize{LEFT: A black hole in AdS that receives a kick at $t_0$. Arbitrary trajectories in AdS can be generated by a dense sequence of such instantaneous kicks, allowing us to describe proper time evolution in any weakly curved spacetime.  RIGHT: Twin black holes. The left twin is static while the right twin is the accelerated black hole of the LEFT panel. The time dilation experienced by the twins is computed by the modular Berry holonomy of the ``loop'' of modular Hamiltonians describing the two trajectories and the integral of the zero mode projection of the CFT Hamiltonian along the loop via eq. (\ref{timedilationloop}), (\ref{timedilation}). }}
\label{fig:acceleratedbh}
\end{center}
\end{figure}

\subsection{Time dilation for twin observers}
\label{subsec:timedilation}
The proper time measured by a bulk observer is a gauge dependent quantity, being a function of the initial and final points between which the proper length of the worldline is computed. This fact was reflected in our previous discussion in the choice of the bulk slices $\Sigma_{t_i}$ and $\Sigma_{t_f}$ on which the atmosphere operators are defined. Waiving the need for the latter requires asking a gauge invariant question.

In this Section, we are interested in computing the relative time, or time dilation, between two twin observers who follow different paths through spacetime until they meet at a later boundary time t. Each observer is described in the CFT by a family of modular Hamiltonians $K_1(t)$ and $K_2(t)$. At their meeting events $t_i=0$ and $t_f=t$, the two black holes are near each other so  their local atmosphere operator sets $S^{1,2}_0$ and $S^{1,2}_t$ are related by simple unitaries $U_{12}(0)$ and $U_{12}(t)$ respectively (fig.~\ref{fig:acceleratedbh}), which we assume known.  

Working in the Schrodinger picture, the operators $S^1_t$ at the final meeting time can be obtained from $S^1_0$ via the map (\ref{atmosphereisomorphism}), in two different ways, depending on whether we propagate them along the worldline of the first or the second twin. The two paths are distinguished quantum mechanically by whether $V_S(0,t)$ in (\ref{VSexp}) is constructed  from the modular Wilson line for the family $K_1(t)$ or from the Wilson line of $K_2(t)$ with the appropriate inclusion of $U_{12}(0)$, $U_{12}(t)$. Equivalence of these two procedures implies that the two modular Wilson lines satisfy:
\myal{V_S^{(1)}(0,t)&= U^\dagger_{12}(t) V_S^{(2)}(0,t) U_{12}(0) \nonumber\\
&\Rightarrow {\cal W}_1(0,t) e^{-i\sum_a c_1(t) Q^1_a(0)}  = U^\dagger_{12}(t) {\cal W}_2(0,t) U_{12}(0) e^{-i\sum_a c_2(t) Q^1_a(0)}\nonumber\\
&\Rightarrow U_{12}^\dagger(0){\cal W}_2^\dagger U_{12}(t) {\cal W}_1 = e^{-i\sum_a c_2(t) Q^1_a(0)} e^{i\sum_a c_1(t) Q^1_a(0)} \label{observerloop}}
where $Q^1_a(0)$ are the zero modes of $K_1(0)$ and, in the second line, we used the fact that $Q^2_a(0) = U_{12}(0) Q^1_a(0) U^\dagger_{12}(0)$. The two families of modular Hamiltonians in this problem, together with the unitaries that relate them at the initial and final moments, form a closed operator ``loop'', therefore, the L.H.S of eq. (\ref{observerloop}) is an example of a modular Berry holonomy ${\cal W}_{\text{loop}}$ discussed in Section \ref{subsec:KandW}.

According to our proposal, each observer's proper time is the coefficient of the modular Hamiltonian in the evolution operators $V^{(1)}_H(0,t)$, $V^{(2)}_H(0,t)$ given by eq. (\ref{atmisomorphismH}). To measure the time dilation between the two observers we have to look at the coefficient of $K$ in the operator $ U_{21}^\dagger (0) V_H^{(2) } U_{21}(0)  V_H^{(1)\dagger} $ which by virtue of (\ref{atmisomorphismH}) and (\ref{observerloop}) becomes:
\myal{ &U_{12}^\dagger (0) V_H^{(2) } U_{12}(0)  V_H^{(1)\dagger}\nonumber\\
&= \exp\left[i\int_0^t dt'\, U_{12}^\dagger(0) \,e^{iHt'} P_0^{(2) t'}[H]e^{-iHt'}\,U_{12}(0)\right] {\cal W}_{\text{loop}}  \exp\left[-i\int_0^t dt'\, e^{iHt'} P_0^{(1) t'}[H]e^{-iHt'}\right] \label{timedilationloop}}
The result (\ref{timedilationloop}) is a unitary operator generated by \emph{modular zero modes} of $K_1(0)$ that depends only on the CFT Hamiltonian and an intrinsic property of the two black holes: the families of modular Hamiltonians $K_1(t)$, $K_2(t)$ describing the time evolution of their state and the relation of their instantaneous frames at their meeting points $U_{12}(0),U_{12}(t)$. As per our proposal in Section \ref{subsec:proposal}, the proper time is identified with the coefficient of the modular Hamiltonian in the modular eigenoperator decomposition of 

\myeq{-i\log \left[U_{12}^\dagger (0) V_H^{(2) }U_{12}(0) V_H^{(1)\dagger}\right] = (2\pi)^{-1}\Delta \tau_{12} \,K_1(0)  +\sum_a c'_a \,Q_a^1(0) \label{timedilation} }

\paragraph{Exercise} The reader is encouraged to use the technology explained in Section \ref{subsec:adsBH} to compute the left hand side of (\ref{timedilation}) for the twin black holes of fig.~\ref{fig:acceleratedbh} and confirm that $\Delta \tau_{12}$ yields the correct time dilation.

\section{Particle detection}
\label{sec:particles}

\begin{figure}
\begin{center}
\includegraphics[width=7.5cm]{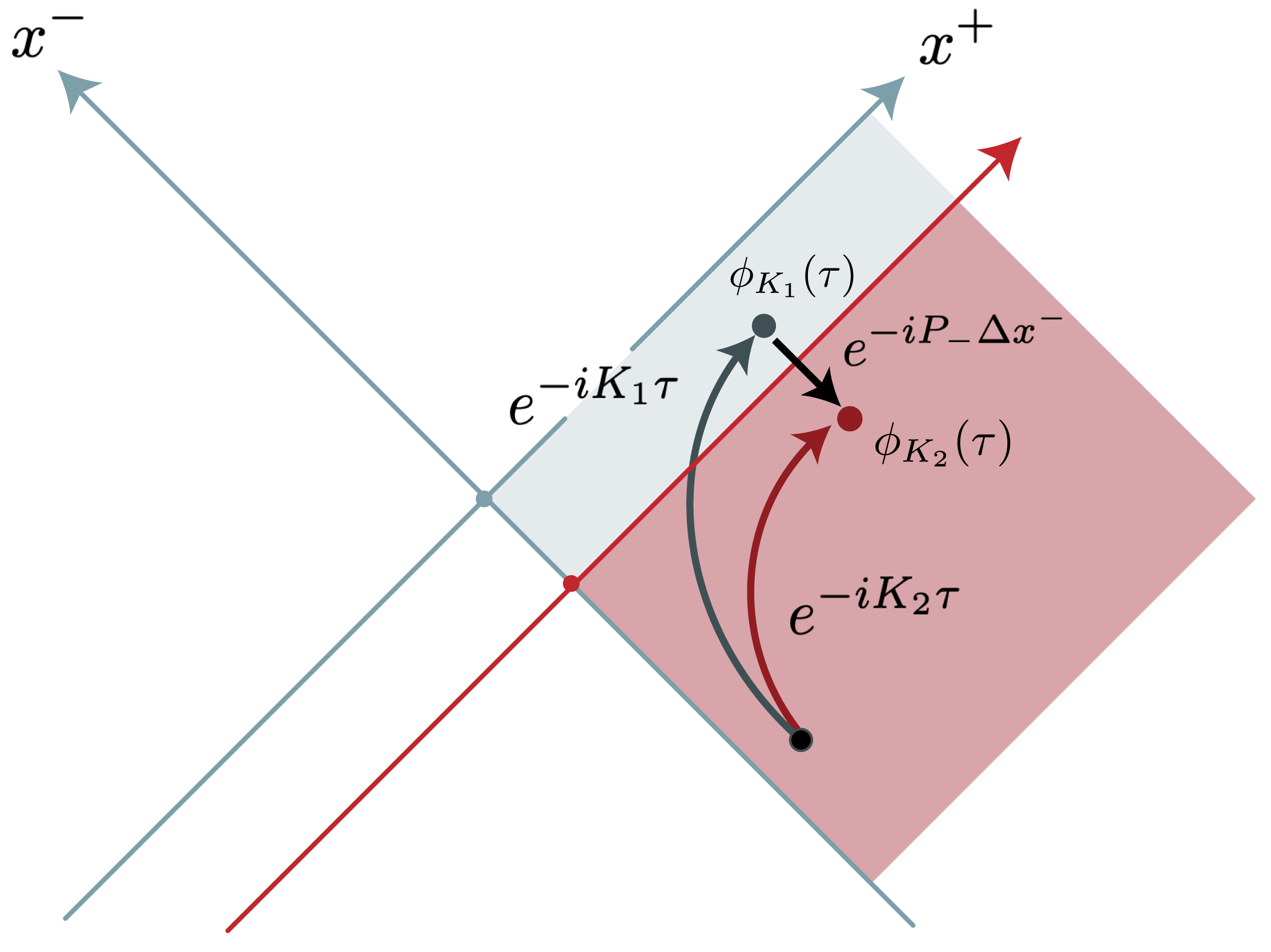}
\includegraphics[width=7.5cm]{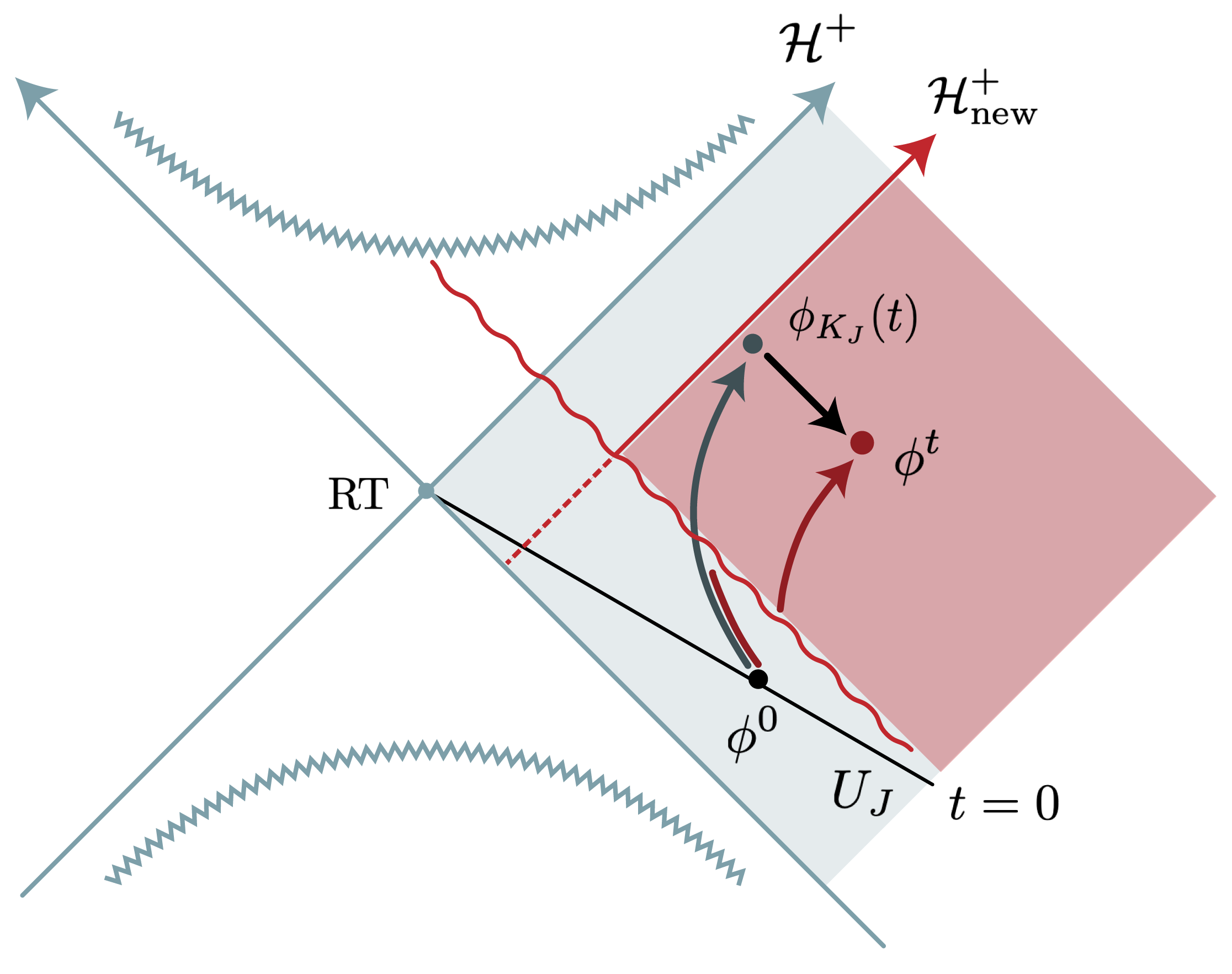}
\caption{\footnotesize{LEFT: Two null separated Rindler wedges in Minkowski space and the corresponding vacuum state modular flows, generating boosts about the boundary of the corresponding wedge. Saturation of modular chaos is manifested in the exponential deviation of the two trajectories. RIGHT: Backreaction of a black hole spacetime due to an infalling particle and the comparison between the action of the modular flowed operator $\phi_{K_J}$ and the proper time evolved operator $\phi^t_H$ on the state $|\Psi_J\rangle$. The former preserves the distance of the excitation from the RT surface whereas the later preservers the distance from the local horizon, up to exponentially decaying corrections. The exponential deviation of the two trajectories at late times reflects the Shapiro shift of the location of the horizon which is manifested quantum mechanically in the saturation of the modular chaos bound---in direct analogy to the physical interpretation of maximal modular chaos in the LEFT panel.  }}\label{modularinclusionfig}
\end{center}
\end{figure}

Up to this point, our black hole was guaranteed an undisturbed journey: no particles were allowed to cross its path. Under this condition, we argued, modular flow of its atmosphere operators amounts to proper time evolution along the worldline of the black hole, in the classical background it lives in. This ceases to be true in the presence of infalling excitations, since the atmosphere is defined relative to the apparent horizon, which becomes shifted (fig. \ref{modularinclusionfig}) with respect to the extremal surface when particles get absorbed.

In this Section we explain that in order to describe proper time evolution of the atmosphere fields, modular flow needs to be corrected by a \emph{modular scrambling mode} $G_{2\pi}$ contribution: an operator that exponentially grows under modular flow $e^{iK\tau}G_{2\pi} e^{-iK\tau}= e^{2\pi \tau} G_{2\pi}$ with an exponent that saturates the modular chaos bound of \cite{deBoer:2019uem, Faulkner:2018faa}.  This physically describes the null shift of the causal horizon of the final black hole relative to the extremal surface. Its coefficient measures the infalling null energy flux at the horizon. This establishes our advertised formula (\ref{claim1}): proper time and infalling energy distribution can be extracted from the unitary relating the initial and final atmosphere operators, by expanding it in the modular eigenoperator basis.

\subsection{Modular flow in the presence of infalling matter}
\label{subsec:Kwithparticles}

Suppose we make a boundary perturbation to a static AdS black hole, so that some particles later fall in. The state of the Universe is then
\myeq{|\Psi_J\rangle= U_J |TFD\rangle = {\cal Z}^{-1/2} \sum_E e^{-\beta E/2} U_J |E\rangle_{\text{sys}} |E\rangle_{\text{ref}} \label{PsiJ}}
where $U_J= e^{-i \sum_i \int J_i(\Omega,r)\phi_i(r,\Omega,t=0)}$ inserts the small perturbation of the supergravity fields $\phi_i$, with $i$ an abstract flavor index, on an initial bulk Cauchy slice $\Sigma_0$. We also assume that the perturbation is introduced far from our probe black hole so that $U_J$ is initially spacelike separated from the ``lab'', the operators within a radius $\ell$ from the black hole
\myeq{ [U_J, \phi^0(\rho, \Omega)]=0, \,\,\,\,\,\text{for:  } 0<\rho < \ell}
The absorption of the perturbative particle, of course, does not affect the proper length of the black hole's worldline at leading order in $1/N$, which in this case coincides with the global time separation of the worldline's endpoints $\tau=\Delta t$. 

In order to understand this example in our formalism, we start by choosing two timeslices $\Sigma_0$ and $\Sigma_t$, where we assume that on $\Sigma_t$ the $U_J$ excitation has already been absorbed by the black hole, namely that it has reached the stretched horizon in Schwarzschild frame. The absorption causes the black hole to grow, resulting in a small perturbation in the near horizon metric at $\Sigma_t$. 

The local atmosphere fields are gravitationally dressed to the \emph{local horizon}, as explained in Section \ref{subsec:codesub}, with time set from the boundary by the slice $\Sigma$. This means that the operator $\phi^0(\rho, \Omega)$ inserts a particle on $\Sigma_0$ at a particular distance $\rho$ from the horizon, when acting on a CFT state dual to the original black hole geometry $|\Psi_J\rangle$ or small fluctuations about it. Since the metric on $\Sigma_t$ is only perturbatively different from that of $\Sigma_0$ (since now the black hole is assumed to remain stationary at the center of AdS), the \emph{Schrodinger picture} atmosphere operators at the final slice $\phi^t$ will be the \emph{same} as $\phi^0$: Acting with $\phi^t(\rho,\Omega)=\phi^0(\rho,\Omega)$ on $e^{-iHt}|\Psi_J\rangle$ introduces an excitation at the same distance $\rho$ from the \emph{new} local horizon. Switching to the Heisenberg picture we then have:
\myal{\phi^t_H(\rho, \Omega)&= e^{iHt} \phi^0(\rho,\Omega) e^{-iHt}  }
Proper time evolution $V_H(0,t)$ is generated by the CFT Hamiltonian in this case. 

According to our proposal, to read off the proper time we need to express $V_H(0,t)$ in terms of modular flow. The modular Hamiltonian for our system, after tracing out the reference, reads:
\myeq{K_J= 2\pi U_J H U_J^\dagger}
and the corresponding evolution of the atmosphere fields gives
\myeq{\phi_{K_J}(t)= e^{\frac{i}{2\pi} K_J t} \phi^0 e^{-\frac{i}{2\pi}{K_J} t} = \begin{cases}  \phi^t_H & \forall \, t: \,\,\,  [\phi^t_H, U_J]=0 \\ 
U_J \phi^t_H U_J^\dagger &  \forall \, t: \,\,\,  [\phi^t_H, U_J]\neq0 \end{cases} \label{phiKexact} }
At sufficiently small $t$ modular and time evolutions coincide, so our prescription works as in Section \ref{subsec:adsBH}. It fails, however, once time evolution inevitably moves $\phi^t_H$ inside the lightcone of $U_J$, after which modular flow and proper time flow of $\phi^0$ differ by $U_J [ \phi^t_H, U_J^\dagger]$. Understanding this commutator is the goal of this Section. At leading order in $N$ there are two contributions of interest: The free field contribution and the Shapiro delays due to the highly blueshifted infalling particles near the horizon. We discuss them in order.

\begin{figure}
\begin{center}
\includegraphics[width=7.5cm]{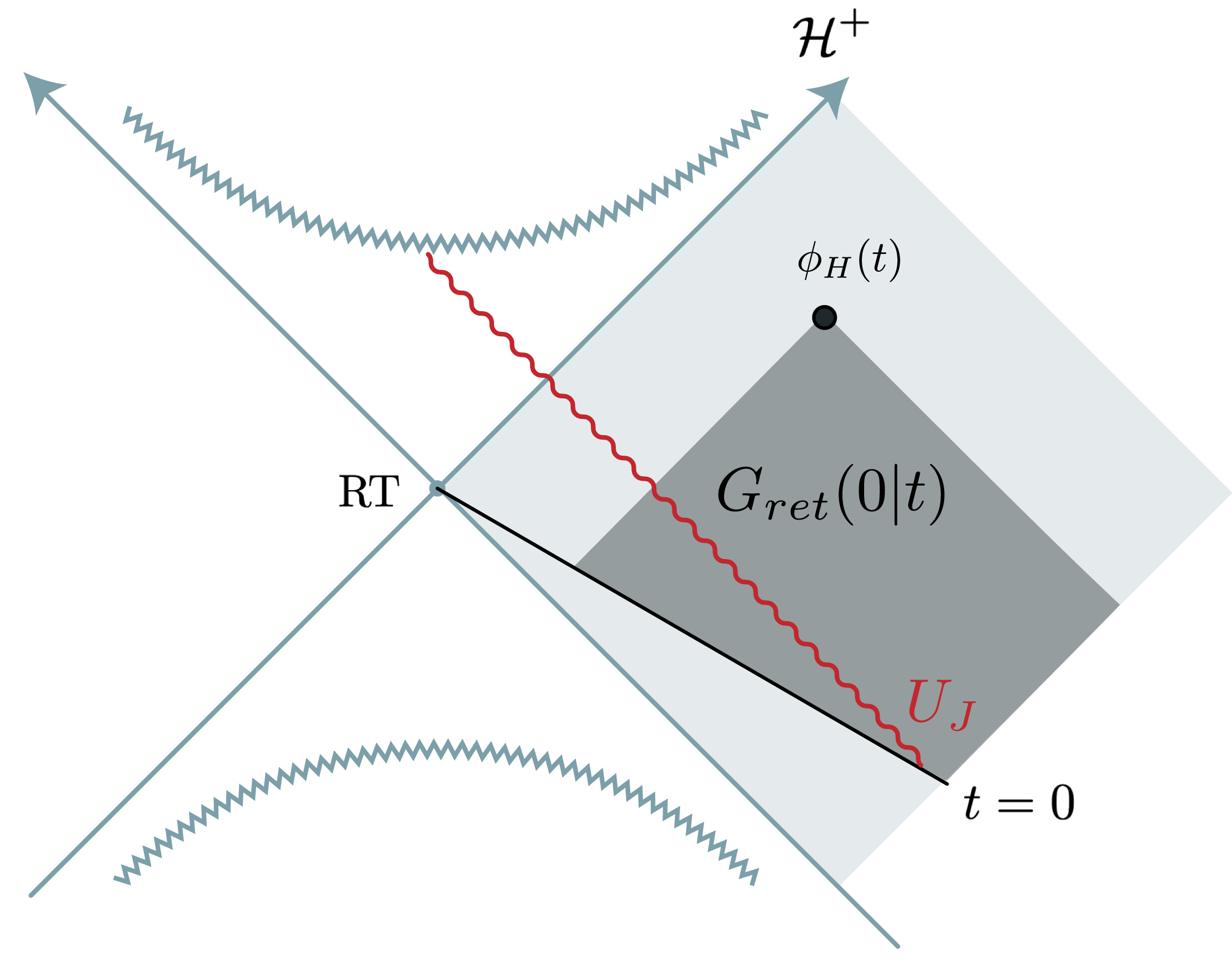}
\includegraphics[width=7.5cm]{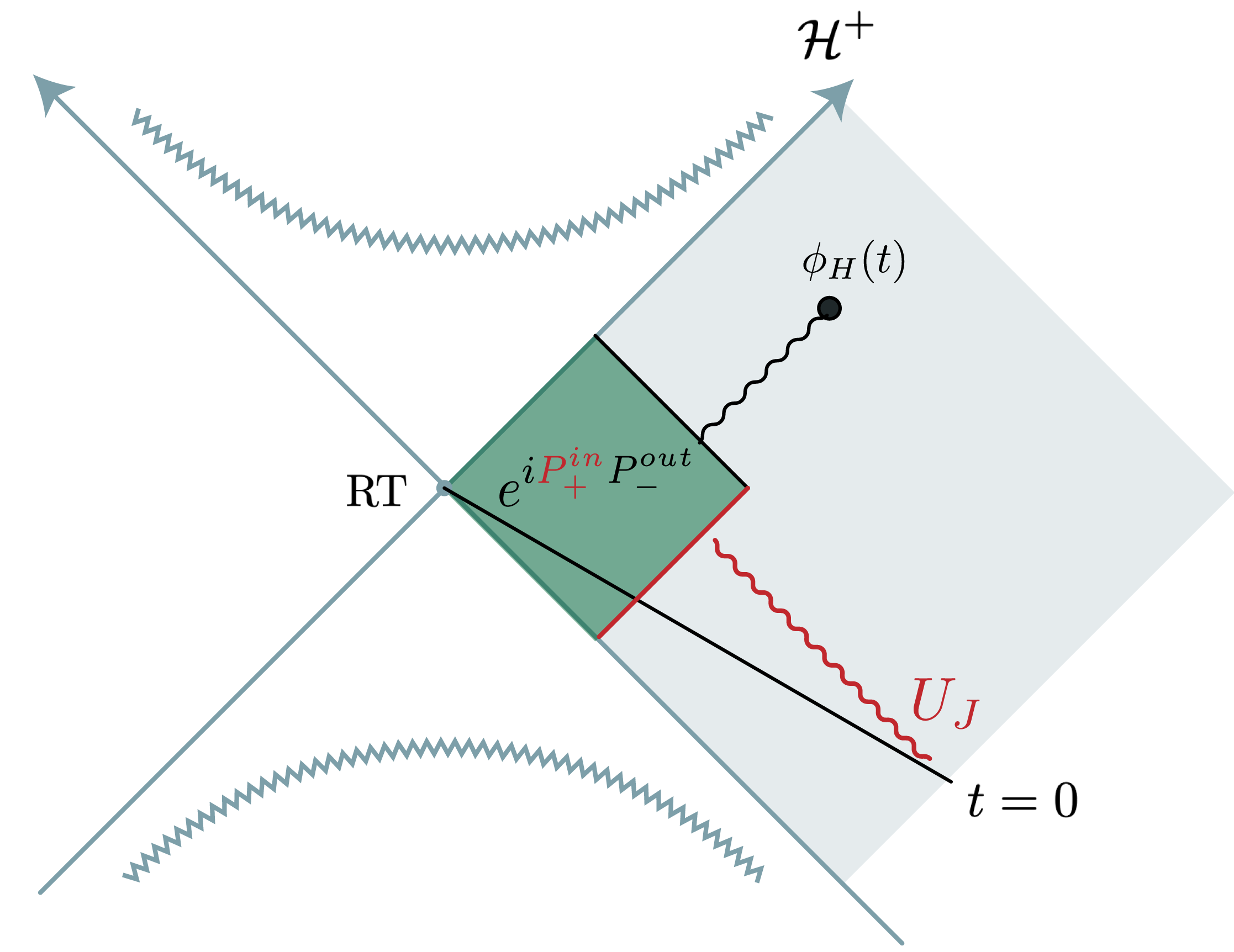}
\caption{\footnotesize{Free field vs shock contributions to the modular flow of a local ``atmosphere'' operator $\phi$ in the state (\ref{PsiJ})}}
\label{freefieldvsshock}
\end{center}
\end{figure}

\paragraph{The free field contribution} At leading order in $N$, the bulk theory is a free QFT on a semi-classical geometry. In this approximation, $\phi^t_H$ inserted at time $t$ can be expressed in terms of $t=0$ fields by usual causal propagation
\myeq{ \phi^t_H(x)=  \int \,dy\, \left(\,\partial_t \,G_{ret}(x, t |y, 0) \phi^0 (y)  + G_{ret}(x,t |y,0 ) \pi^0 (y)\right) }
where $(\phi, \pi)$ a symplectic pair of QFT degrees of freedom. The commutator of interest, in the free field approximation, becomes:
\myal{U_J \left[ \phi^t_H(x), \,U^\dagger_J\right] \Big|_{free}&=-i\int\, dy\, G_{ret}(x,t| y,0)\,U_J \frac{\delta}{\delta \phi^0(y)} U^\dagger_J =  \int \,dy \, G_{ret}(x,t | y,0) \,J(y) \\
&=-\langle \Psi_J | \phi_H^t(x) |\Psi_J\rangle \label{freefieldK}}
In contrast to the geometric proper time evolution, modular flow removes the expectation value that $\phi$ acquires in the reference state. This is a version of the ``frozen vacuum'' problem, inherent in many entanglement based approaches to bulk reconstruction. The operators of interest to us are located near a black hole horizon so the relevant $G_{ret}$ is controlled by the quasi-normal modes and decays exponentially in proper time
\myeq{\langle  \Psi_J | \phi_H^t |\Psi_J\rangle \sim e^{- t} }
after crossing the future lightcone of $U_J$. With the assumption that our chosen final moment is at least a few thermal times later than the last infalling quantum, we can safely neglect this contribution to modular flow.

It is, of course, possible to consider more general bulk QFT excitations, for example: 
\myeq{U'_J= \exp\left[ \frac{i}{2} \int  dx_1 dx_2\,J(x_1,x_2)\, \phi^0(x_1) \,\phi^0(x_2) 
+\dots\right]}
The free field contribution (\ref{freefieldK}) follows from the same reasoning and yields the non-local operator 
\myeq{U'_J \left[ \phi^t_H(x), \,U^{'\dagger}_J\right] \Big|_{free}= \int dy_1 dy_2\, G_{ret}(x,t | y_1,0)J(y_1,y_2) \phi^0 (y_2) +\dots}
The important observation is that, once again, these operator contributions are exponentially decaying in time, after crossing the lightcone of $U'_J$, due to the retarded propagator contribution to the smearing function
\myeq{U'_J \left[ \phi^t_H(x), \,U^{'\dagger}_J\right] \Big|_{free}\sim e^{- t}}
These non-local contributions, therefore, also become negligible, by assuming enough proper time separation between $\Sigma_t$ and the last infalling particle.

\paragraph{The shock contribution} In the absence of a black hole, the free field result (\ref{freefieldK}) would have been the dominant contribution to the commutator, since all higher order corrections coming from interactions would be suppressed by powers of $1/N$. The large redshift of the near horizon metric, however, accelerates the infalling quantum exponentially as it approaches $r_{BH}$. This exponential increase of its energy in the local Schwarzschild frame competes with the $G$ suppression of gravitational interactions and results in a non-trivial change in the propagation of the atmosphere operators \cite{Shenker:2013pqa}.

This gravitational intuition is reflected quantum mechanically in the observation that the overlap of the state $\phi_{K_J}(t,r, \Omega)|\Psi_J\rangle $ with $\phi^t_H(r,\Omega) |\Psi_J\rangle$ is, in this case, equal to a familiar, out-of-time-order bulk correlation function \cite{Maldacena:2015waa} in the thermofield double state:
\myeq{\langle\Psi_J| \,\phi_H^t(r,\Omega) \,\phi_{K_J}(t,r,\Omega) |\Psi_J\rangle= \langle U_J^\dagger \,\phi^t_H(r,\Omega) \,U_J\, \phi _H^t (r,\Omega) \,\rangle_{TFD} \label{BHotoc}}
In theories of gravity, for $t$ sufficiently large, the scattering of $\phi_H^t$ and the infalling particle $U_J$ takes place very close to the horizon. Due to the near horizon geometry, the infalling particle's null energy is exponentially blueshifted in the frame of the particle $\phi_H^t$, $\langle P_+\rangle = e^{t} \,\delta E^0_+$ where $\delta E^0_+ \sim O(1)$ is the null energy of $U_J$ in the $t=0$ frame, when the excitation was introduced. The effect of such a blueshifted infalling particle on the propagation of $\phi^t_H$ can be approximated by a null shockwave with some spatial distribution along the transverse directions $\Omega$, which results in a null translation of $\phi_H$ \cite{tHooft:1996rdg, Kiem:1995iy}:
\myal{\langle\Psi_J| \phi^t_H\, \phi_{K_J}(t) |\Psi_J\rangle &\approx \langle \phi^t_H\exp \left[-i\int d\Omega \,\Delta x^- (t,\Omega) P_-(\Omega) \right] \phi^t_H \rangle_{TFD} \label{shockOTOC} \\
\text{where: }\Delta x^-(t,\Omega)&= \int d\Omega'\, f(\Omega, \Omega')\, \langle \,U_J^\dagger \,P_+ (\Omega')\, U_J\,\rangle_{TFD} \label{Shapiro}\\
 P_\pm (\Omega)&= \int dx^\pm \,\,T^{\text{bulk}}_{\pm\pm} (\Omega, x^\mp=0)}
$\Delta x^-(t,\Omega)$ is the Shapiro time delay caused by the infalling $U_J$ which grows as $e^{ t}$ for $1\ll t\ll \log N$, and the smearing function $G(\Omega, \Omega')$ is a transverse propagator along the horizon, satisfying $(\nabla^2_{\Omega} -1)f(\Omega, \Omega') =-2\pi \delta (\Omega, \Omega')$. We have assumed here that the perturbation $U_J$ results in a semi-classical spacetime, so that   $\Delta x^-$ can be replaced by its expectation value at leading order.

The exponentially growing Shapiro delay results in the exponential decay of the overlap (\ref{shockOTOC}) and the states $\phi_{K_J}(t)|\Psi_J\rangle$, $\phi^t_H|\Psi_J\rangle$ become nearly orthogonal after the scrambling time. This implies that modular evolution is not a good approximation to the geometric proper time evolution when there is infalling energy. Nevertheless, eq.~(\ref{shockOTOC}) shows how to fix this. Consider the operators $G_{2\pi}(\Omega)=U_J \,P_-(\Omega)\, U^\dagger_J$ which obey:
\myeq{[K_J, \,G_{2\pi}(\Omega)]=-2\pi i \, G_{2\pi}(\Omega) \label{scramblingmode}}
$G_{2\pi}$ were called modular scrambling modes in \cite{deBoer:2019uem} and are discussed further in Section \ref{subsec:modchaos}. It straightforwardly follows from (\ref{shockOTOC}) that 
\myeq{\langle\Psi_J| \phi^t_H\,e^{i\int\,d\Omega\, \Delta x^- (t, \Omega) \,G_{2\pi}(\Omega)} \phi_{K_J}(t)\,e^{-i\int\,d\Omega\, \Delta x^- (t, \Omega) \,G_{2\pi}(\Omega)} |\Psi_J\rangle  \approx 1 \label{scramblingcorrection}}
assuming an appropriate smearing of the local atmosphere operator $\phi$ so that the state $\phi |\Psi_J\rangle$ is normalized to $1$.

\paragraph{The result} Our observation (\ref{scramblingcorrection}) illustrates that proper time evolution of the ``lab'' degrees of freedom $\phi^0$ continues to be related to modular flow, at leading order in $N$, but the two no longer coincide; modular evolution needs to supplemented by scrambling mode contributions to account for the infalling particle's backreaction on the relative location of the atmosphere and the extremal surface:
\myeq{  \phi^t_H\approx \begin{cases}  \phi_{K_J}(t) & \forall \, t: \,\,\,  [\phi^t_H, U_J]=0 \\ 
e^{i\int d\Omega \Delta x^-(t,\Omega) G_{2\pi}(\Omega)} \phi_{K_J}(t) e^{-i\int d\Omega \Delta x^- (t, \Omega) G_{2\pi}(\Omega)}    + O\left(e^{- t},\frac{1}{N}\right) &  \forall \, t: \,\,\,  [\phi^t_H, U_J]\neq0 \end{cases} \label{phiKshock} }

We can now combine the scrambling mode and modular flows above, using the Baker-Campbell-Hausdorff relation, the commutator (\ref{scramblingmode}), the Shapiro delay (\ref{Shapiro}) and the fact that  $\langle \Psi_J|  P_+ (\Omega)|\Psi_J \rangle =\delta E^0_+ (\Omega) e^{t}$ where $\delta E^0_+(\Omega)$ is the local averaged null energy at the horizon in the frame of the $t=0$ timeslice $\Sigma_0$, to obtain
\myal{\phi^t_H&= V_H(0,t) \,\phi^0 \,V^\dagger_H(0,t)\nonumber\\
V_H(0,t)&\approx \exp\left[ \frac{i}{2\pi} \int  \,\delta E^0_+ (\Omega')  \, f(\Omega, \Omega') \,G_{2\pi}(\Omega)\,t  +\frac{i}{2\pi}K_J \,t + O\left(e^{-t}, N^{-1}\right)  \right] \label{fullproposal}}
Eq. (\ref{fullproposal}) is an example of our general claim (\ref{claim1}) advertised in the introduction: Proper time evolution along the worldline of our black hole $V_H$ can be organized in terms of operators of definite modular weight, with the coefficient of the modular Hamiltonian measuring proper time and the coefficient of the scrambling mode $G_{2\pi}$ measuring the infalling null energy distribution at the horizon.

\paragraph{Moving Black Holes.}
The generalization of the result (\ref{fullproposal}) to the Section \ref{subsec:adsBH} scenario of black holes  in a general semi-classical asymptotically AdS spacetime  is straightforward. For example, starting with the state for the boosted black hole in empty AdS and exciting infalling bulk QFT modes as before we get
\myal{|\Psi_{J,\eta}\rangle &= U_J e^{-iB\eta}|TFD\rangle \label{movingandparticle}}
where $B$ is the generator of the boost symmetry of AdS, giving our black hole rapidity $\eta$, while $U_J= e^{-i \sum_i \int J_i(\Omega,r)\phi_i(r,\Omega,t=0)}$. 

The CFT representation of the Heisenberg picture atmosphere operators on the initial ($t_i=0$) and final ($t_f=t$) Cauchy slices are given by (\ref{boostedphi0}) (\ref{boostedphit}), with $\phi_{static}$ being the HKLL formula for a local bulk field in a static black hole background, with a perturbative gravitational dressing to the local horizon:
\myal{\phi^0_H &= e^{-iB\eta} \phi_{static}(r,\Omega) e^{iB\eta} \label{earlyphi}\\
\phi^t_H & = e^{iHt} e^{-iP x(t)} e^{-iB\eta(t)} \phi_{static}(r,\Omega) e^{iB\eta(t)}e^{iPx(t)} e^{-iHt} \label{latephi}}
The functions $x(t), \eta(t), \tau(t)$ describe the location, momentum and proper time of the black hole in the AdS background. 

Following the previous reasoning, it can be shown that the proper time evolution can be expressed in terms of a flow generated by modular eigenoperators as:
\myal{V_H(0,t) &= \exp\left[ \frac{i}{2\pi} \int \delta E^0_+(\Omega) \,f(\Omega, \Omega') \,G_{2\pi}(\Omega') \,\tau(t) +\frac{i}{2\pi} K_{J,\eta} \,\tau(t) +O\left(e^{-t},N^{-1}\right) \right]  \label{fullproposal2}}

where: \myal{ K_{J, \eta}& = 2\pi \,U_J \,e^{-iB\eta} \,H\, e^{iB\eta} \,U^\dagger_J\nonumber\\
\delta E^0_+(\Omega) &= \langle \Psi_{J,\eta}| e^{-iB\eta}\,P_+(\Omega) \,e^{iB\eta}| \Psi_{J,\eta} \rangle \nonumber\\
\tau(t) &= \tan^{-1}\frac{\tan t}{\cosh\eta}\nonumber\\
[K_{J,\eta}\,,\,G_{2\pi}] &= -2\pi i G_{2\pi}
}
This is another illustration of our main claim, where $\tau(t)$ is the proper length of the black hole's trajectory and $\delta E_+^0 (\Omega)$ is the average null energy crossing the causal horizon at angle $\Omega$ in the frame of the initial $\Sigma_0$ bulk timeslice.

\subsection{Energy distribution from maximal modular chaos}
\label{subsec:modchaos}

Let us now place the results of Section \ref{subsec:Kwithparticles} within a more general framework. It was argued in \cite{deBoer:2019uem, Faulkner:2018faa} that analyticity properties of general QFT modular Hamiltonians, $K$, imply an upper bound on the modular weight of $\delta K$, where the latter denotes the first order perturbation of $K$ due to a state excitation or an infinitesimal change of the subalgebra of interest. This bound on modular chaos can be articulated as the condition:
\myeq{ \lim_{1\ll |\tau| \ll\log N} \Big| \frac{d}{d\tau} \log | \langle O_2| e^{iK_\Psi\tau}\,\delta K \,e^{-iK_\Psi\tau}|O_1\rangle| \Big| \leq 2\pi\, ,\,\,\,\, \,\, \forall\,\, |O\rangle = O |\Psi\rangle \label{chaosbound}}
where the $O$ operators  act within the bulk code subspace. The modular scrambling modes are  operators that saturate this bound as $\tau \to \pm \infty$ and can be extracted from $\delta K$ formally as:
\myeq{ G_{\pm} = \pm\frac{1}{2\pi } \lim_{\tau\to \pm \infty} e^{-2\pi |\tau |} e^{iK_{\Psi} \tau}\, \delta K\,e^{-iK_{\Psi} \tau} \label{GfromdK}}
Note that the limit here should be taken after the large $N$ limit, so that $\tau$ remains less than or of order the scrambling time.

The prototypical example of maximal modular chaos involves two Rindler wedges in Minkowski space, with their entanglement surfaces being separated by a null deformation (fig. \ref{modularinclusionfig}). The modular Hamiltonians for two wedges in the vacuum state, which equal Rindler boost generators about the two entanglement surfaces, form the Poincare algebra \cite{Faulkner:2016mzt, Casini:2017roe}:
\myeq{[K_1, K_2] = 2\pi i (K_1-K_2) \label{inclusion}}
Eq. (\ref{inclusion}) continues to hold for arbitrary null deformations of the Rindler wedge and the corresponding subalgebras are said to form a \emph{modular inclusion}. The operator $G_{2\pi}=K_2-K_1$ saturates the bound (\ref{chaosbound}) and can be shown to generate a location dependent null shift at the entangling surface. 

Maximal modular chaos, therefore, reflects the geometric structure of the QFT background: Saturation of (\ref{chaosbound}) for $\tau \to \pm \infty$ is a diagnostic of the inclusion properties of spatial subalgebras, and the corresponding scrambling modes (\ref{GfromdK}) encode the local Poincare algebra near the region's edge. This motivated \cite{deBoer:2019uem} to propose the use of modular chaos in holography, where the structure bulk spacetime is not \emph{a priori} known, as a principle for extracting the local Poincare algebra, directly from the CFT. The results of the previous Section can be understood in this framework, as we now explain.

\paragraph{Maximal modular chaos from infalling particles} 
The two protagonists of this paper have been the CFT modular flow, $e^{iK\tau}$, and the proper time evolution of the atmosphere fields along the black hole worldline, $V_H$. In absence of infalling energy in the state $|\Psi\rangle$ the two were argued to coincide. Infalling particles whose backreaction away from the probe black hole can be neglected are included by acting with a unitary $W$, so that our state $|\Psi\rangle = W |\tilde\Psi \rangle$ where $|\tilde\Psi \rangle$ describes the same spacetime without any particles that fall into the probe, the proper time evolution is given by the modular Hamiltonian $\tilde{K}$ associated to $|\tilde\Psi\rangle$. The bound (\ref{chaosbound}), therefore, applies to the difference between modular and proper time Hamiltonians. 

Our results (\ref{fullproposal}), (\ref{fullproposal2}) show that a state excitation that introduces an amount of infalling energy flux through ${\cal H}^+$, leads to a modular Hamiltonian perturbation that saturates (\ref{chaosbound}). This guarantees that no operators with higher modular weight can appear in the modular eigenoperator expansion of $\log V_H$. The bound (\ref{chaosbound})  is saturated in theories in which the bulk dual is Einstein gravity, in the sense that there is a large higher spin gap. Then the Averaged Null Energy distribution at the horizon can, then, be extracted from $\log V_H$ by taking the limit
\myeq{\frac{1}{2\pi}\int d\Omega d\Omega'\, \delta E_+^0(\Omega') \,f(\Omega',\Omega)  \, G_{2\pi}(\Omega)\, \tau(t) =\lim_{s\to + \infty}  e^{-2\pi |s|} e^{iK s} i\log V_H(0,t) e^{-iK s} \label{extractingdE}}
Conversely, the vanishing of the R.H.S. of eq. (\ref{extractingdE}) signifies that no particles crossed the horizon. 

The physical interpretation of (\ref{extractingdE}) is very analogous to the Rindler example of maximal modular chaos above. The original modular flow, $e^{iKt}$, continues  to boost the atmosphere fields about the Ryu-Takayanagi surface even past the shockwave and (at least) until the scrambling time $\tau \lesssim \log S_{BH}$ ---up to exponentially decaying corrections--- whereas proper time evolution $V_H=e^{i\tilde{K}t}$ preserves the location of the fields in the local AdS-Schwarzschild frame which gets non-trivially shifted in the null direction after crossing the shock (fig. \ref{modularinclusionfig}). The geometric action of the two modular flows, $K$ and $\tilde{K}$, is reminiscent of the case of included algebras in flat space and the saturation of the bound (\ref{chaosbound}) is a manifestation of this inclusion property \cite{Jefferson:2018ksk}, with $G_{2\pi}$ implementing the relevant null shift. The additional feature of the present case is that the null separation of the two ``included wedges'' is given by the, appropriately smeared, null energy of the absorbed particles.

\section{Discussion}
\label{sec:discussion}

\subsection{Resolution of the Marolf-Wall puzzle}
\label{subsec:MarolfWall}

We may now return to the opening question of Section \ref{sec:intro}: The experience of an observer falling into a black hole which we will take to be an eternal, two sided AdS black hole. This bulk configuration is described holographically by two decoupled conformal CFT$_L \times$ CFT$_R$ in the highly entangled, thermofield double state. 

As first emphasized by Marolf and Wall \cite{Marolf:2012xe} in the early days of the firewall debates, this setup presents us with a conceptual puzzle: Entanglement wedge reconstruction allows us to introduce an observer somewhere in the right black hole exterior by acting only with CFT$_R$ operators. Since the two CFTs are decoupled, we are guaranteed that the observer is composed by CFT$_R$ degrees of freedom for the entire \emph{boundary time} evolution and, thus \emph{commutes} with all CFT$_L$ operators. On the other hand, ER=EPR suggests that the bulk dual to $|TFD\rangle$ is an Einstein-Rosen bridge with a smooth interior geometry. The bulk observer's trajectory crosses the right black hole horizon at finite \emph{proper time} and after horizon crossing, the observer can receive signals sent from the left exterior which implies that its degrees of freedom \emph{do not commute} with the CFT$_L$ operator algebra. Proper time evolution must, therefore, couple the two CFTs, despite the absence of a microscopic dynamical coupling! This seemingly bizarre conclusion appears to suggest that either the two decoupled CFTs in $|TFD\rangle$ cannot predict the experience of the infalling observer beyond the horizon without further specifying some coupling between the two sides \cite{Gao:2016bin, Kourkoulou:2017zaj}, or that the $|TFD\rangle$ does not actually describe a connected geometry and our observer's experience can be reconstructed entirely from CFT$_R$, while their detection of particles coming from the left is merely a mirage.

\begin{figure}
\begin{center}
\includegraphics[width=8cm]{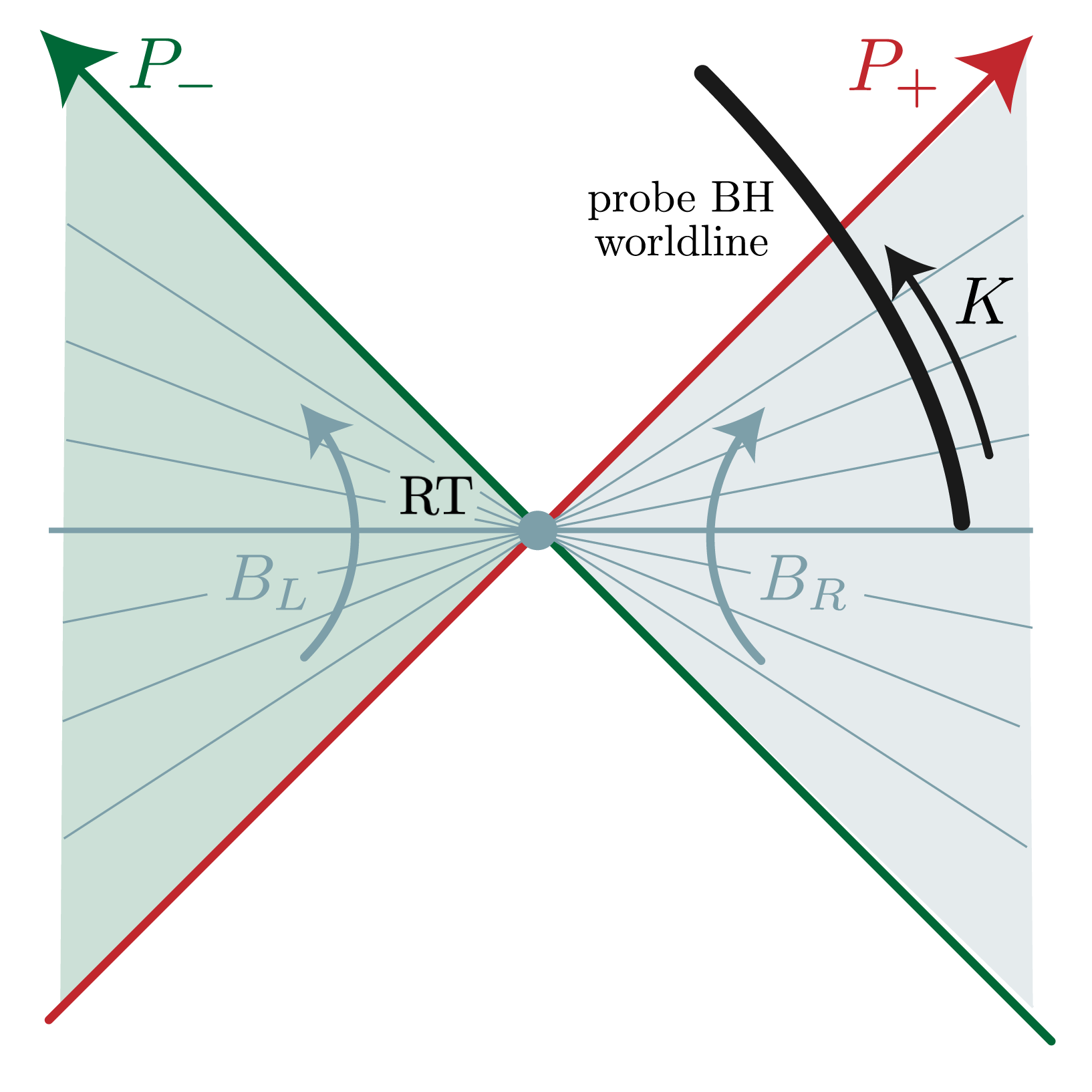}
\caption{\footnotesize{The modular Hamiltonian $K$ of the probe black hole propagates the local atmosphere operators in proper time along its worldline, even past the Rindler horizon. $K$ can be expressed as a linear combination (\ref{Klinearcomb}) of the two (decoupled) Rindler Hamiltonians $B_L$ and $B_R$ which preserve the left and right wedge, respectively, and the ANEC operators $P_\pm$ that shift the RT surface along null directions and, therefore, mix the two operator algebras, resolving the Marolf-Wall puzzle. }}
\label{fig:marolfwall}
\end{center}
\end{figure}

The puzzle is resolved quite elegantly in our framework. The right bulk observer is introduced in our setup by thermally entangling a subset of the CFT$_R$ degrees of freedom with an external reference and collapsing them into a black hole, somewhere near the right asymptotic boundary. Our proposal says that, as long as nothing falls in our probe black hole, proper time evolution of the atmosphere operators, even past the horizon, is generated by the Left-Right system's modular Hamiltonian, obtained by tracing out the reference. Insofar as $\rho_{RL}= \text{Tr}_{ref}\big[ |\psi\rangle_{L,R,ref}\,_{L,R,ref}\langle \psi|\big]$ is not Left-Right separable, it is clear that modular flow $\rho^{i\tau}_{LR}$ will generically mix the Left and Right algebras, thus naturally evading the puzzle. In other words, for states that are sufficiently entangled to describe a short bulk wormhole connecting the two exteriors, entangling the reference with degrees of freedom of CFT$_R$, necessarily entangles it with CFT$_L$ as well, and the observer's modular flow couples the two sides, allowing proper time evolution to access the common interior. 

For a simple illustration of the resolution, consider the Rindler decomposition of Anti-de Sitter spacetime, where we introduce our probe black hole inside the Right wedge and near the asymptotic boundary $\rho\to \infty$, far away from the Rindler horizon (fig. \ref{fig:marolfwall}). Upon tracing out the reference system, the modular Hamiltonian of our system at $t=0$ reads:
\myeq{K(0) = e^{-iP\rho } H e^{iP\rho} \sim \frac{e^{\rho}}{2} \left(  B_L- B_R + P_+ + P_-\right) \label{Klinearcomb} }
$B_L$ and $B_R$ are the Rindler Hamiltonians of the left and right Rindler wedges, respectively. These generate automorphisms of the corresponding algebras and, consequently, do not mix the two sides of the hyperbolic black hole. In contrast, $P_+$ and $P_-$ are the ANE operators along the two Rindler horizons which are related to the global Hamlitonian $H$ and AdS translation isometry $P$ by $H= P_+ + P_- $ and $P= P_+ - P_-$. They generate null shifts of the bifurcation surface, resulting in a flow that mixes the left and right algebras. 

In particular, evolution of the atmosphere operators $\phi$ in proper time would, according to our formalism, correspond to modular flow:
\myal{\phi_K (\tau)& = e^{iK\tau} \phi e^{-iK\tau} = e^{iP_- e^{\tau} } e^{iB\tau } \phi e^{-iB\tau} e^{-iP_- e^\tau} +O(e^{-\tau})\nonumber\\
&= e^{iP_- e^{\tau} }  \phi_B(\tau) e^{-iP_- e^\tau} +O(e^{-\tau})}
The key thing to observe is the appearance of the exponentially growing null shift $P_-$ which will translate the Rindler evolved field $\phi_B(\tau)$ past the Rindler horizon after a \emph{finite} proper time $\tau$.

By analogy, we hypothesize that, in the background of an eternal black hole, evolving operators in the right asymptotic region by the modular Hamiltonian of an infalling observer introduced in CFT$_R$, as in fig.~\ref{fig:marolfwall}, schematically reads, at sufficiently late proper time $\tau$:
\myeq{\phi_K(\tau) = e^{iK\tau} \phi e^{-iK\tau} \sim e^{iP_- e^\tau} \phi_H(\tau) e^{-iP_- e^\tau} +O(e^{-\tau}) \label{infallingbh}}
where the exponentially growing, ANE operator contributions  $P_\pm$ to the observer's modular Hamiltonian $K= -\log \rho_{LR}$ will appear in the form of left/right operator products $O_L O_R$, as in \cite{Lin:2019qwu}. Such products are expected to appear due to the entanglement of the two CFTs. Preliminary calculations of the modular Hamiltonian of an infalling observer in an SYK setup similar to \cite{Gao:2019nyj} confirm this hypothesis for AdS$_2$ black holes, where the ANE operator contributions to $K$ appear in the form of the size operator of \cite{Qi:2018bje}.

\subsection{The frozen vacuum problem}
\label{subsec:frozenvac}
Our prescription, as outlined above, is an explicit method for reconstructing the operators in the black hole interior. Importantly, it is also extremely simple, utilizing only the CFT dual of the atmosphere operators on the initial slice and the modular Hamiltonian $K=-\log \rho_{LR}$. Its relation to the Papadodimas-Raju proposal for the black hole interior \cite{Papadodimas:2012aq} will be discussed in related upcoming work \cite{dBLL}. 

However, this prescription, as it currently stands, suffers from the ``frozen vacuum'' problem \cite{Bousso:2013ifa}: We can use the modular flowed operators (\ref{infallingbh}) to create particles in the black hole interior, or detect excitations of the initial state we used to construct the modular Hamiltonian, but we cannot measure excitations already present in the initial state. This is a consequence of the fact that modular evolution of an operator preserves its expectation value in the given state and is therefore blind to the causal effect that originally spacelike separated excitations can have on the operator at later times. 

In our particular setup, we were able to evade the frozen vacuum problem by assuming knowledge of the local atmosphere operators on the final timeslice. The comparison of the initial and final operators in the CFT allowed us to reconstruct our black hole's ``history'', including whether it encountered any energy on its path. Such knowledge of the final operators, however, can not reasonably be assumed for an observer that falls inside a black hole. Resolution of the frozen vacuum in this case seems to require some new conceptual element.

\subsection{How small can our probe black hole be?}
In the main body of this work we considered fairly large probe black holes, with horizon radii of the order of the AdS scale $L_{AdS}$. These black holes are simplest to describe because they dominate the canonical ensemble and, therefore, thermally entangling them with a reference is described by the ``canonical'' thermofield double (\ref{genstate}) between the system and the reference CFTs. The resulting system modular Hamiltonian, then, reads $K= U_{sys} H_{CFT} U_{sys}^\dagger$ which has a relatively simple action. The price we pay with this simplification is that we can only probe features of the AdS universe at cosmological scales.

In order to probe the bulk geometry at sub-AdS scales, we need smaller black holes. Black holes with $R_H < L_{AdS}$ cannot be described by the canonical ensemble, due to their negative specific heat. There exists, however, a \emph{parametrically} large window of smaller than $L_{AdS}$ black holes that dominate the \emph{microcanonical} ensemble \cite{Horowitz:1999uv}. This can be seen by a back-of-the-envelope calculation. The thermodynamic behavior of small black holes in $AdS$ can be approximated by that of their flat space cousins. A ${d+1}-$dimensional black hole with energy $E$ such that $R_H \sim \ell_{pl}^{\frac{d-1}{d-2}} E^{\frac{1}{d-2}} <L_{AdS}$ has an entropy:
\myeq{S_{BH} \sim  \left( \frac{R_H}{\ell_{pl}}\right)^{d-1} \sim \ell_{pl} (\ell_{pl} E)^{\frac{d-1}{d-2}}}
On the other hand, the competing configuration, a thermal gas of supergravity excitations of the same total energy, has an entropy that scales like a gas of massless particles in a box of size $L_{AdS}$\footnote{The same formula also applies when an internal manifold is present, assuming its size is $O(L_{AdS})$. The only difference is that the box along the internal manifold directions has periodic ---instead of reflective--- boundary conditions. } :
\myeq{S_{gas} \sim (L_{AdS}E)^{\frac{d}{d+1}}}
The two configurations exchange dominance when $S_{BH}\sim S_{gas}$ which happens at energy $E\sim \ell^{-1}_{pl} \left( \frac{L_{AdS}}{\ell_{pl}}\right)^{\frac{d(d-2)}{2d-1}}$, when the black hole radius reaches:
\myeq{ R_{H} \sim \ell_{pl} \left( \frac{L_{AdS}}{\ell_{pl}}\right)^{\frac{d}{2d-1}}
}
The important observation is that small black holes entropically dominate over a thermal gas of the same energy, for horizon radii that are parametrically smaller than $L_{AdS}$ for any dimension $d>1$ as can be seen by the $L_{AdS}/\ell_{pl} \to \infty $ limit of the ratio
\myeq{ \frac{R_H}{L_{AdS}} \sim  \left(\frac{L_{AdS}}{\ell_{pl}}\right)^{\frac{1-d}{2d-1}} \to 0}
To get a sense of how small these black holes can get, consider the case of AdS$_4$ with curvature radius comparable to the Hubble length $L_{AdS} \sim 10^{26} \,\rm{m}$ and Planck length $\ell_{pl} \sim 10^{-35}\,\rm{m}$. The smallest microcanonically stable black hole has a Schwarzschild radius $R_H \sim 100\,\rm{m}$, comparable to the size of a physics department!

As explained in more detail in \cite{Horowitz:1999uv}, the microcanonical equilibrium states in the energy window 
\myeq{ \ell^{-1}_{pl} \left( \frac{L_{AdS}}{\ell_{pl}}\right)^{\frac{d(d-2)}{2d-1}} \lesssim E \lesssim \ell_{pl}^{-1} \left( \frac{L_{AdS}}{\ell_{pl}} \right)^{d-2}} 
should be understood as a coexistence phase between small black holes and thermal gas, with most of the total energy stored in the black hole. Due to its negative specific heat, a small black hole in AdS will initially radiate some of its energy but the entropic argument above suggests that it will quickly equilibrate with its thermal atmosphere, as long as we keep the total energy fixed.

\paragraph{``Microcanonical'' thermofield double} The previous thermodynamic argument suggests that small probe black holes thermally entangled with a reference can be described quantum mechanically by the microcanonical version of the thermofield double state \cite{Marolf:2018ldl}:
\myeq{ |TFD\rangle_{micro} = {\cal Z}^{-{1/2}} \sum_n e^{-b E_{n}} f(E_{n}) |E_n\rangle_{sys} |E_n\rangle_{ref} \label{microtfd}}
where $f(E)$ a smooth function of energy that effectively projects the coherent sum onto a microcanonical window of width $\sigma$ about a fixed energy $E_0$. A simple example of such a function is a Gaussian $f(E)\propto \exp \left[ -(E -E_0)^2/\sigma^2 \right]$. Note that the coefficient $b>0$ in (\ref{microtfd}) is a free parameter, not to be confused with the inverse temperature which is microcanonically \emph{defined} via $\beta= \frac{\partial S}{\partial E}$.

The gravitational duals of the microcanonical wormholes (\ref{microtfd}) were studied in detail in \cite{Marolf:2018ldl}, where it was shown that the bulk Euclidean path integral preparation of this state is dominated by a semi-classical saddle configuration describing a small black hole, as long as the width of the energy window satisfies:
\myeq{1\ll \sigma \ll G_N^{-1/2} \label{window}}
For energy windows that are too narrow, $\sigma \lesssim O(1)$ the uncertainty principle implies large quantum fluctuations in the relative time of the two exteriors $\Delta t > O(1)$ so the clocks at the two ends of the wormhole are decohered, and the state does not describe a semi-classical wormhole. On the other hand, a wide window effectively takes us back to the canonical ensemble and our small black hole becomes unstable.

In order to introduce a small ``black hole observer'' in a general spacetime we therefore simply have to replace (\ref{genstate}) with the analogous unitary excitation of (\ref{microtfd}):
\myeq{|\Psi\rangle=  {\cal Z}^{-{1/2}} \sum_n e^{-b E_{n}} f(E_{n}) U_{sys} |E_n\rangle_{sys} |E_n\rangle_{ref} \label{microgen}}

\paragraph{Code subspace modular Hamiltonian} The central ingredient in our proposal for describing the proper time propagation of the atmosphere fields was the modular Hamiltonian of the system obtained after tracing the reference. More specifically, we only cared about its projection onto the bulk code subspace $S_0$, roughly consisting of excitations with $O(1)$ energy about the background state (\ref{microgen}). Given that the function $f(E)$ is approximately constant within an energy window that can scale with $N$ (\ref{window}), $K=-\log \rho_{sys}$ on the code subspace acts simply as a unitary rotation of dynamical CFT Hamiltonian up to $G_N$ corrections
\myeq{K \propto U_{sys}\,H\,U_{sys}^\dagger + O(G_N). \label{microK}}
If the state (\ref{microgen}) is described by a dual semiclassical black hole geometry with a timelike killing vector near the black hole horizon, the modular Hamiltonian (\ref{microK}) will act geometrically within the black hole atmosphere, up to the corrections from infalling particles discussed in Section \ref{sec:particles}. In this case, the reasoning of Section \ref{sec:time} goes through, extending the validity of our prescription to small observers and offering a useful tool for exploring sub-AdS locality.

However, there is a subtlety with the assumption that the reduced state obtained from (\ref{microgen}) describes a semiclassical black hole. The problem can be seen by recalling that localized wavepackets in flat space spread out in time in a diffusive fashion $\Delta r \sim \sqrt{\frac{t}{m}}$, where $m$ is the particle's mass. Similarly, the wavefunction of an initially localized small black hole will tend to spread over an $L_{AdS}$ sized region in time. The microcanonical equilibrium state obtained from (\ref{microgen}) by tracing out the reference, therefore, does not describe a single classical geometry but rather an ensemble of macroscopically distinct spacetimes with the black hole located at different points within an $L_{AdS}$ region. Notice that this is not an issue for the canonical black holes with $R_{H}\geq L_{AdS}$ because the gravitational potential preserves the localization of the wavepacket. 

In order to construct a classical bulk observer, therefore, the simple state (\ref{microgen}) does not suffice: We need to further ``measure'' the black hole location, i.e. project the state onto a localized wavepacket. This could conceivably be done by performing the corresponding  measurement on the reference side, where the black hole lives in an empty AdS Universe and the localization can be achieved by exploiting the AdS isometries. The resulting state will be only in approximate equilibrium with the corrections set by the rate of the wavepacket spreading. We leave a detailed exploration of this interesting construction of sub-AdS scale observers for future study.

\subsection{Emergent time}
\label{subsec:emergenttime}

\begin{figure}
\begin{center}
\includegraphics[width=10cm]{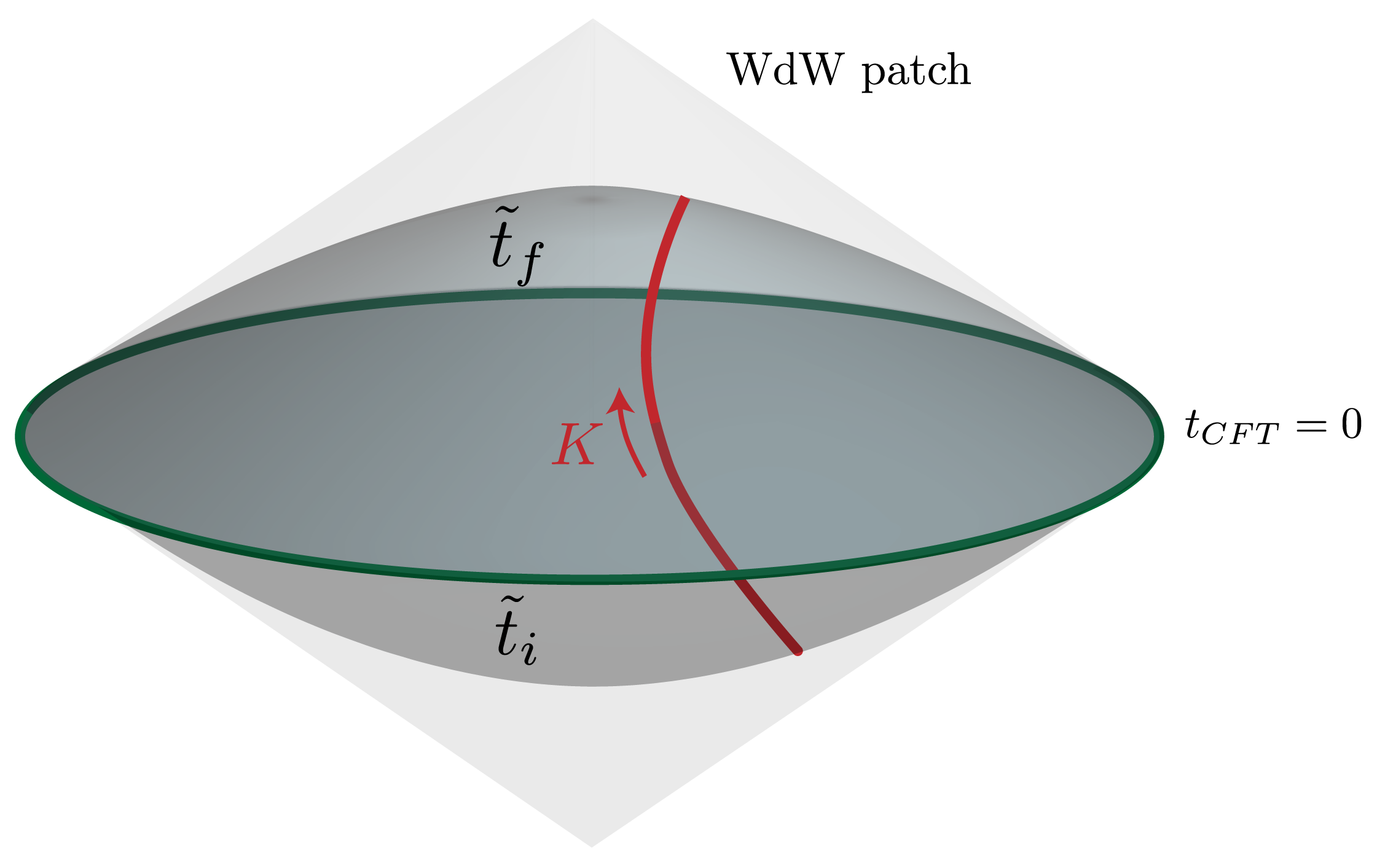}
\caption{\footnotesize{A Wheeler-de Witt patch in AdS corresponding to the CFT state at a fixed boundary time $t_{CFT}=0$. Our observer (red worldline) travels between the two Cauchy slices of the WdW patch labelled by $\tilde{t}_{i,f}$. The proper length of the worldline can be computed via eq. (\ref{claim1}), even though no dynamical evolution of the full quantum system takes place. }}
\label{fig:WdW}
\end{center}
\end{figure}

Our proposal serves as a step towards demystifying the internal notion of time in holographic, gravitational systems. The central idea is simple: The observer is a physical system, entangled with the world. Tracing out the observer, endows the rest of the system with a modular Hamiltonian which defines the time flow in their reference frame, insofar as the observer remains undisturbed ---and with corrections of the type discussed in Section \ref{sec:particles} for perturbative disturbances. By its very construction, this is an inherently \emph{relational clock} that becomes available due to the quantum entanglement between the observer and the environment, adding one more entry to the growing list of gravitational concepts whose roots can be traced to ubiquitous features of quantum systems \cite{Ryu:2006bv, Gao:2016bin, Maldacena:2013xja, Susskind:2014rva, Susskind:2017ney, Susskind:2018tei, Brown:2019hmk, Czech:2018kvg}. The importance of entanglement between the clock and its environment, and of the modular automorphism in particular, in the emergence of time has been discussed in the past, \cite{Page:1983uc, Connes:1994hv}. Our work descends from the same conceptual lineage.

Crucially, our ``proper time Hamiltonian'' (\ref{claim1}) does not rely on the existence of any global notion of time evolution. In our AdS example, the initial and final timeslices $\Sigma_{\tilde{t}_i}, \Sigma_{\tilde{t}_f}$ could be chosen to be Cauchy slices in the same Wheeler-de Witt patch, asymptoting to the same CFT time (fig.~\ref{fig:WdW}). The construction of the proper time evolution $V_H(\tilde{t}_i,\tilde{t}_f)$ would follow identical steps to those of Sections \ref{sec:time} and \ref{sec:particles}, with the only difference that the contribution from the zero mode projection of the CFT Hamiltonian in eq. (\ref{atmisomorphismH}) would be absent. In addition, we made very limited use of the asymptotic AdS boundary. We may, therefore, be optimistic that our approach can serve as the seed for a more general framework of emergent time in cosmological quantum gravity models.

\acknowledgments{LL is grateful to Jan de Boer for collaboration on related topics and numerous illuminating conversations and to Bartek Czech for discussions and feedback on the draft. We would also like to thank Raphael Bousso, Tom Faulkner, Juan Maldacena, Jamie Sully, Lenny Susskind, Erik Verlinde, Herman Verlinde. for discussions. DLJ and LL both acknowledge the hospitality of the Aspen Center for Theoretical Physics, the ``Amsterdam String Workshop 2019'' and the KITP program ``Gravitational Holography'', where parts of this work were completed. DLJ is supported in part by DOE Award DE-SC0019219. LL is supported by the Pappalardo Fellowship. }

\end{document}